\documentclass[twocolumn,aps,prb,superscriptaddress]{revtex4-2}

\usepackage{mathtools}
\usepackage{amssymb}
\usepackage{subcaption}
\usepackage{ragged2e}
\usepackage[hidelinks]{hyperref}
\usepackage{color}
\usepackage{tikz}
\usetikzlibrary{arrows.meta,calc,decorations.pathreplacing,positioning}
\newcommand{\capj}[1]{#1}

\newcommand{\inb}[3]{{\left#1 #2 \right#3}}
\newcommand{\reff}[1]{{Fig.\ref{fig:#1}}}
\newcommand{\refs}[1]{{Sec.\ref{sec:#1}}}
\newcommand{\refe}[1]{{Eq.(\ref{eq:#1})}}
\newcommand{\refee}[2]{{Eqs.(\ref{eq:#1}) and (\ref{eq:#2})}}
\newcommand{\refeee}[2]{{Eqs.(\ref{eq:#1}-\ref{eq:#2})}}
\newcommand{\squad}{\ \ }

\begin{document}

\title{ 
Nucleation of Sachdev-Ye-Kitaev Clusters in One Spatial Dimension}

\author{Hrant Topchyan}
\affiliation{A. Alikhanyan National Science Laboratory, Yerevan 0036, Armenia}

\author{Tigran A. Sedrakyan}
\affiliation{A. Alikhanyan National Science Laboratory, Yerevan 0036, Armenia}
\affiliation{Max-Planck-Institut f{\"u}r Physik komplexer Systeme, N\"othnitzer Str. 38, 01187 Dresden, Germany}
\affiliation{Department of Physics, University of Massachusetts, Amherst, Massachusetts 01003, USA}

\begin{abstract}
We study how Sachdev-Ye-Kitaev (SYK) interactions can arise from localized single-particle states on a system that is effectively one dimensional. 
If a local interaction is projected onto coarse localized orbitals, the resulting couplings do not immediately follow the standard SYK distribution. 
Instead, they have a finite probability of being exactly zero, a broad non-Gaussian distribution for the nonzero values, and strong correlations coming from the geometry of the localized states.
We then show that this changes when each localization volume is resolved into $M>1$ smaller microscopic pieces with random phases. 
As $M$ increases, the distribution of the nonzero couplings moves toward the complex-Gaussian SYK form. 
At the same time, 
the large-$M$ limit is a sparse but asymptotically canonical SYK network: the nonzero couplings create SYK clusters, while the pattern of missing or very weak couplings is still determined by the real-space overlap of the localized orbitals.
Finally, we map the interaction tensor to a graph in pair space. 
This makes it possible to follow the formation, merger, and growth of SYK clusters, which we characterize using connected components and clique/simplex counts. The result is a minimal real-space phenomenological theory of SYK-cluster formation, providing clear experimental criteria.

\end{abstract}

\maketitle

\section{Introduction}

The Sachdev-Ye-Kitaev (SYK) model is a setting in which strong interactions, random all-to-all couplings, non-Fermi-liquid dynamics, and many-body quantum chaos can be studied on equal footing. In the large-$N$ limit it is governed by melonic diagrams, develops an emergent infrared conformal regime, and provides a controlled bridge between strongly correlated quantum matter, strange-metal phenomenology, and nearly-AdS$_2$ holography~\cite{SachdevYe1993,Kitaev2015,Sachdev2015,MaldacenaStanford2016,MaldacenaStanfordYang2016,KitaevSuh2018,ChowdhuryGeorgesParcolletSachdev2022,Gu2020}. At the same time, the finite-$N$ model has become a testing ground for spectral statistics, ergodicity, transport, sparsity, and nonperturbative field-theoretic methods~\cite{GarciaGarciaVerbaarschot2016,GarciaGarciaVerbaarschot2017,GarciaGarciaJiaVerbaarschot2018,GarciaGarciaJiaRosaVerbaarschot2021,AltlandBagretsKamenev2017,AltlandBagrets2018,AltlandBagretsKamenev2019Transport,AltlandBagretsKamenev2019Granular,SedrakyanEfetov2017,SedrakyanEfetov2020}. This broader perspective is especially important for condensed-matter realizations, where one must distinguish a Hamiltonian that is merely SYK-like in selected observables from a microscopic construction that actually reproduces the SYK disorder law.

In this paper we focus on the complex-fermion $q=4$ SYK model, whose Hamiltonian is

\begin{equation}
\begin{split}
    H_{\text{SYK}} &= -\mu \sum_i c_i^\dagger c_i +
    \sum_{i,j,k,l} \bar{J}_{ij}^{kl} c_i^\dagger c_j^\dagger c_k c_l =\\&=
    -\mu \sum_i c_i^\dagger c_i +
    \sum_{i<j, k<l} J_{ij}^{kl} c_i^\dagger c_j^\dagger c_k c_l
    \label{eq:syk}
\end{split}
\end{equation}
where the four-fermion coefficients
$J_{ij}^{kl}=\bar{J}_{ij}^{kl}-\bar{J}_{ji}^{kl}-\bar{J}_{ij}^{lk}+\bar{J}_{ji}^{lk}$
are independent (up to antisymmetry and hermiticity constraints) complex Gaussian random variables in the canonical SYK ensemble, $c_i$ and $c_i^\dagger$ are canonical fermionic annihilation and creation operators defined in an orthonormal basis, and $\mu$ is the chemical potential.
For complex fermions, hermiticity requires $\bar J_{ij}^{kl}=\bar J_{kl}^{ij\,*}$, while fermionic statistics imply antisymmetry under $i\leftrightarrow j$ and $k\leftrightarrow l$, i.e.
$J_{ij}^{kl}=-J_{ji}^{kl}=-J_{ij}^{lk}$.
These constraints are implemented explicitly by the antisymmetrization above and by restricting the sums to $i<j$ and $k<l$ in the second line of Eq.~\ref{eq:syk}.
The defining SYK ingredient is therefore not randomness by itself but a very specific disorder law: after the symmetry constraints are imposed, the independent couplings are zero-mean complex Gaussian variables with variance $\langle |J_{ij}^{kl}|^2\rangle\propto N^{-3}$.
That law is what guarantees the familiar large-$N$ limit and the corresponding infrared physics.

In a condensed matter system with spatially local interactions, the effective quartic couplings are obtained by projecting the underlying two-body interaction onto single-particle wave functions.
As a result, the same orbital centers, widths, envelopes, and phases enter many different matrix elements at the same time.
The disorder in the couplings is therefore not arbitrary.
Different couplings are correlated, many of them are exactly zero because the relevant real-space overlaps vanish, 
and only in special limits can the nonzero part of the distribution approach the circular complex-Gaussian law of the canonical SYK ensemble.
The central question of this paper is to determine when that SYK limit can emerge. This general point is familiar from one-body models with correlated disorder, where correlations in the random variables can generate critical or delocalized states at special energies and thereby qualitatively modify localization and transport \cite{XieDasSarma1988,Flores1989,DunlapWuPhillips1990,PhillipsWuScience1991,WuPhillips1991,Sedrakyan2004RDM,SedrakyanOssipov2004,SedrakyanKestnerDasSarma2011}. Although the present setting is different, since the disorder here lives in the interaction tensor rather than in a one-body potential, the same lesson applies: disorder correlations can be essential for the resulting observables.

This question is directly motivated by current efforts to realize SYK physics in experimental systems.
Early solid-state proposals pointed out that sufficiently irregular mesoscopic structures can produce SYK-type interactions~\cite{PikulinFranz2017,LantagneHurtubiseLiFranz2018,ChewEssinAlicea2017,DanshitaHanadaTezuka2017}.
For complex fermions, an especially important setting is graphene flakes in a strong magnetic field, where zeroth-Landau-level states become sensitive to the sample boundary and spatially localized.
In that case, an irregular geometry can drive the projected interaction tensor toward the SYK ensemble~\cite{ChenIlanDeJuanPikulinFranz2018,CanNicaFranz2019,BrzezinskaGuanYazyevSachdevKruchkov2023}.
Similar ideas also arise in flat-band and moir\'e systems.
For example, a disordered optical kagome lattice was shown to generate complex-SYK couplings by projecting interactions onto disorder-selected flat-band states~\cite{WeiSedrakyan2021}, while disordered magic-angle twisted bilayer graphene was argued to form weakly coupled SYK bundles and to exhibit strange-metal behavior~\cite{WeiSedrakyan2023}.
More generally, compact localized states and isolated flat bands, including artificial flat-band platforms, geometrically frustrated lattices, and disorder-structured kagome realizations, provide exactly the ingredients that matter here: spatially bounded orbitals, strong sensitivity to geometry, and a natural route to effectively low-dimensional interaction networks~\cite{LeykamAndreanovFlach2018,DoraKailasvuoriMoessner2011,BilitewskiMoessner2018,MaksymenkoMoessnerShtengel2017,MaitiSedrakyan2019,BaeSedrakyanMaiti2023,LeeAndreanovSedrakyanFlach2024}.

A useful microscopic complement to this picture is provided by the analysis of interaction effects in a smooth random magnetic field by one of us and Raikh~\cite{SedrakyanRaikh2008}. We showed that even a nonquantizing random field can curve semiclassical trajectories, generate additional spatially varying phases in the electronic propagators, and introduce a new field-dependent low-energy structure in the density of states. Applied to the present problem, this means that a smooth magnetic texture can help produce exactly the kind of complex internal phase structure that is needed below for Gaussianization of the active couplings. At the same time, it can generate an additional non-SYK spectral background. Thus, in some magnetic quantum materials, the emergence of SYK clusters, and thus the strange metal physics, can coexist with a field-induced phase texture and possibly with extra low-energy spectral features.

Recent work has studied many related directions, including bosonic holographic analogs, electron-phonon and Yukawa-SYK routes to non-Fermi liquids and unconventional superconductivity~\cite{SwingleWiner2024,EsterlisSchmalian2019,ValentinisInkofSchmalian2023,CichutekRueckriegelKopietz2024,LiValentinisPatelGuoSchmalianSachdevEsterlis2024}, transport and thermalization, sparse and non-Hermitian variants, Brownian spectral probes of chaos, graph-based routes to many-body quantum chaos, and dissipative topological extensions~\cite{JhaKehreinLouw2025,ChapmanDemulderGalanteSheoreyShoval2025,NandyPathakTezuka2025,KalsiRomitoSchomerus2025,AndreanovCarregaMuruganOlleRosaShir2025,GarciaGarciaSaVerbaarschotYin2025,JaramilloJhaKehrein2025,MohamadiAbouie2025}.
Our goal is different but complementary.
Instead of starting from a structured SYK Hamiltonian and then studying its dynamics, we begin with localized one-particle states in real space and ask how both the SYK disorder law and the associated interaction graph emerge from microscopic geometry.

The paper has two main results. The first is a one-dimensional route to the complex-SYK Gaussian law on the continuous nonzero part of the coupling distribution, which we will call the \emph{active sector}. If each orbital is represented by a single coarse support interval, then the couplings are rotationally invariant in the complex plane but remain strongly non-Gaussian, with a finite weight at zero and a broad regular sector set by overlap geometry. We then show that this distribution Gaussianizes when each localization volume is resolved into many microscopic pieces with random phases. We study two closely related constructions: a random partition ensemble and its equal-cell limit. In both cases the same large-$M$ mechanism operates: the active sector becomes complex Gaussian, correlations between distinct typical couplings are suppressed, and the overlap matrix approaches the identity, so the orthogonalization needed for canonical fermions is controlled. 
The pattern of vanishing couplings is still determined by which localized orbitals overlap in real space. Therefore, even in the large-$M$ limit, the system does not become a single fully connected SYK dot. Instead, it approaches a sparse interaction tensor whose nonzero couplings acquire the canonical SYK statistics (disjoint SYK clusters).

The second main result is a graph-based description of how SYK clusters form. We represent the interaction tensor $J_{ij}^{kl}$ as a weighted graph in pair space. In this graph, each vertex corresponds to an unordered pair of orbitals $(i,j)$, and each edge represents the strength of scattering between two such pairs. If we keep only the strong edges, the connected components of the resulting graph define emergent SYK clusters.
This graph picture is useful because it gives a clear way to measure how close a cluster is to all-to-all SYK mixing. In graph language, an $n$-simplex is a set of $n+1$ vertices that are all connected to one another: a $1$-simplex is an edge, a $2$-simplex is a triangle, and a $3$-simplex is a tetrahedron. In our problem, these simplexes count groups of mutually connected scattering channels and therefore show how close a cluster is to dense mixing in pair space.

These results also clarify what kind of single-particle system our phenomenology is meant to describe. 
The partition ensembles should be viewed as effective models for localized orbitals living on a geometry that is effectively one dimensional, such as a rough boundary, an edge, a filamentary structure, or a multichannel quasi-one-dimensional system. 
They are not meant to represent exact eigenstates of a strictly one-channel real 1D Schr\"odinger Hamiltonian. 
In such a problem, the eigenfunctions can be chosen real, so the independently phased microscopic pieces used in our construction would not be present. 
A more natural microscopic setting is a fermionic multicomponent or higher-dimensional Hamiltonian that localizes the overall envelope while allowing the internal wave function to be complex, for example, because of a magnetic field, channel mixing, or complex hopping. In such a setting, spatial overlap between localized orbitals is not a problem. 
The essential requirement for fermions is that the canonical single-particle orbitals are orthonormal, and they do not need to have disjoint support.

The experimental implication is straightforward. To realize this physics, one needs localized fermionic modes on a structure that is effectively one dimensional, sufficient spatial overlap between nearby modes, and a local interaction. One also needs enough internal phase randomness so that a typical nonzero matrix element is built from many small contributions with unrelated phases. When these conditions are satisfied, the nonzero couplings become Gaussian at large $M$. The remaining pattern of zero couplings then organizes the system into clusters, which can be analyzed using graph theory. In this sense, the present work provides a minimal phenomenological description of how SYK clusters can emerge in one dimension.

The rest of the paper is organized as follows.
In Section~\ref{sec:correlated_disorder}, we develop the exact continuum formulation and the correlated-disorder statistics generated by localized orbitals.
In Section~\ref{sec:true_syk} we show how internal phase structure turns the active sector into the Gaussian SYK clusters and makes the basis asymptotically canonical.
Finally, Section~\ref{sec:graph_mapping} recasts the same microscopic data as an interaction graph and quantifies the nucleation, growth, and simplex scaling of the emergent SYK clusters.

\section{Correlated-disorder route to SYK couplings}
\label{sec:correlated_disorder}

We begin with the minimal continuum problem that underlies the rest of the paper: project a spatially local two-body interaction onto a random basis of localized single-particle orbitals in one dimension. The resulting four-fermion coefficients $J_{ij}^{kl}$ are therefore not independent and identically distributed (IID) disorder variables. They inherit the same geometric disorder from the underlying orbital data. Two consequences follow immediately. Different couplings are mutually correlated because they depend on shared centers, widths, and phases, and a finite fraction vanish identically whenever the required real-space overlap intervals are empty. This is the physically natural baseline from which the true SYK limit must emerge. It is also closely related in spirit to other structured SYK constructions discussed in the literature~\cite{LantagneHurtubiseLiFranz2018}. We will build directly on this baseline, asking respectively how the disorder law can be Gaussianized and how the residual sparsity organizes the couplings into emergent clusters.

\subsection{Geometric construction and disorder ensemble}
\label{sec:geometric_construction}

We consider a one-dimensional system of length $L$ with open boundary conditions. Each localized orbital is modeled by a rectangular wave function characterized by a center $c_i$, a width $w_i$, and a phase $\varphi_i$,
\begin{equation}
    \psi_i(r) = \frac{e^{i \varphi_i}}{\sqrt{w_i}} \theta\inb({w_i - 2|r - c_i|})
    \label{eq:eigens}
\end{equation}
where $\theta(x)$ is the Heaviside function. With this convention each orbital is individually normalized, $\int_0^{L}\!dr\,|\psi_i(r)|^2=1$ when $S_i\subset[0,L]$. The boundary corrections are $O(w_i/L)$ and are negligible for $\overline{w}\ll 1$.

Here Eq.~\ref{eq:eigens} should be understood as a coarse-grained representation of a localized single-particle state. It retains only the ingredients that matter for projected interaction matrix elements: a localization center, a spatial extent, and a rapidly fluctuating internal phase. The sharp rectangular profile is chosen because it makes the overlap geometry completely explicit and allows closed-form formulas. Replacing the step function with a smooth localized envelope would not change the two mechanisms that drive the discussion below: strict selection rules from empty overlaps and a random U(1) phase multiplying the induced couplings.

The phases and centers are sampled independently and uniformly, $\varphi_i\in[0,2\pi)$ and $c_i\in[0,L]$. For the widths we study two ensembles: a constant-width ensemble $w_i=\overline{w}L$ and a variable-width ensemble with $w_i$ uniform on $[0,\overline{w}L]$. Here $\overline{w}\in(0,1)$ controls the typical localization scale. Throughout, we focus on $\overline{w}\ll 1$, where orbitals are parametrically shorter than the sample and edge corrections at $r=0,L$ are negligible. The total number of orbitals is denoted by $N$.

Physically, $c_i$ is the random localization center, while $w_i$ is a proxy for the localization length or spatial support of the state. The constant-width and variable-width ensembles, respectively, model a relatively homogeneous set of localized states and a strongly inhomogeneous set with broad size fluctuations. The random phases $\varphi_i$ mimic rapidly varying microscopic signs or phases and are responsible for the rotational invariance of the coupling distribution in the complex plane. 
Eq.~(\ref{eq:eigens}) therefore defines normalized seed envelopes, not yet the final canonical orbitals. Its role is to isolate, in the simplest possible form, how random geometry and a local interaction generate effective SYK couplings. The overall interaction scale is set by the strength of $V(r)$ and may be rescaled globally via $V\to gV$ if desired.

\begin{figure*}[t]
    \centering
    \tikzset{
        stateFill/.style={fill=black!10},
        stateLine/.style={draw=black,thick},
        overlapFill/.style={fill=blue!35,fill opacity=0.35},
        guide/.style={densely dashed,line width=0.6pt},
        dim/.style={line width=0.8pt},
        lab/.style={fill=white,fill opacity=0.92,text opacity=1,inner sep=2.0pt,rounded corners=1.2pt},
        labS/.style={lab,font=\scriptsize},
    }

    \begin{subfigure}[t]{0.49\textwidth}
        \centering
        \begin{tikzpicture}[x=0.78cm,y=0.98cm,>=Latex, every node/.style={font=\footnotesize}]
            \draw[->,line width=0.8pt] (0,0) -- (10.35,0) node[right] {$r$};
            \node[labS,anchor=north,yshift=-2pt] at (0,0) {$0$};

            \fill[stateFill] (1.0,1.70) rectangle (7.0,2.35);
            \fill[stateFill] (4.5,0.45) rectangle (9.0,1.10);

            \fill[overlapFill] (4.5,1.70) rectangle (7.0,2.35);
            \fill[overlapFill] (4.5,0.45) rectangle (7.0,1.10);

            \draw[stateLine] (1.0,1.70) rectangle (7.0,2.35);
            \draw[stateLine] (4.5,0.45) rectangle (9.0,1.10);

            \node[lab] at (2.35,2.03) {$S_i$};
            \node[lab] at (8.05,0.78) {$S_j$};

            \node[labS] at (5.75,1.30) {$S_{ij}=S_i\cap S_j$};

            \draw[guide] (4.5,0.30) -- (4.5,2.55);
            \draw[guide] (7.0,0.30) -- (7.0,2.55);
            \node[labS,anchor=south east,xshift=-10pt,yshift=3pt] at (4.5,2.55) {$S_{ij}^-$};
            \node[labS,anchor=south west,xshift= 10pt,yshift=3pt] at (7.0,2.55) {$S_{ij}^+$};

            \draw[<->,dim] (4.5,2.90) -- (7.0,2.90)
                node[midway,labS,above=3pt] {$\ell_{ij}=|S_{ij}|$};

            \fill (4.0,2.025) circle (1.5pt);
            \fill (6.75,0.775) circle (1.5pt);
            \node[labS,anchor=south,yshift=2pt] at (4.0,2.10) {$c_i$};
            \node[labS,anchor=north,yshift=-2pt] at (6.75,0.70) {$c_j$};

            \draw[decorate,decoration={brace,amplitude=4.5pt}] (1.0,3.30) -- (7.0,3.30);
            \node[labS,anchor=south,yshift=2pt] at (4.0,3.30) {$w_i$};

            \draw[decorate,decoration={brace,mirror,amplitude=4.5pt}] (4.5,-0.65) -- (9.0,-0.65);
            \node[labS,anchor=north,yshift=-2pt] at (6.75,-0.65) {$w_j$};
        \end{tikzpicture}
        \caption{\capj{Finite overlap: $0<\ell_{ij}<\min(w_i,w_j)$.}}
    \end{subfigure}
    \hfill
    \begin{subfigure}[t]{0.49\textwidth}
        \centering
        \begin{tikzpicture}[x=0.78cm,y=0.98cm,>=Latex, every node/.style={font=\footnotesize}]
            \draw[->,line width=0.8pt] (0,0) -- (10.35,0) node[right] {$r$};
            \node[labS,anchor=north,yshift=-2pt] at (0,0) {$0$};

            \fill[stateFill] (1.0,1.70) rectangle (9.2,2.35);
            \fill[stateFill] (3.4,0.45) rectangle (6.0,1.10);

            \fill[overlapFill] (3.4,1.70) rectangle (6.0,2.35);
            \fill[overlapFill] (3.4,0.45) rectangle (6.0,1.10);

            \draw[stateLine] (1.0,1.70) rectangle (9.2,2.35);
            \draw[stateLine] (3.4,0.45) rectangle (6.0,1.10);

            \node[lab] at (7.55,2.03) {$S_i$};
            \node[lab] at (5.45,0.78) {$S_j$};

            \node[labS] at (4.70,1.30) {$S_{ij}=S_j\subset S_i$};

            \draw[guide] (3.4,0.30) -- (3.4,2.55);
            \draw[guide] (6.0,0.30) -- (6.0,2.55);
            \node[labS,anchor=south east,xshift=-10pt,yshift=3pt] at (3.4,2.55) {$S_{ij}^-$};
            \node[labS,anchor=south west,xshift= 10pt,yshift=3pt] at (6.0,2.55) {$S_{ij}^+$};

            \draw[<->,dim] (3.4,2.90) -- (6.0,2.90)
                node[midway,labS,above=3pt] {$\ell_{ij}=w_j$};

            \fill (5.10,2.025) circle (1.5pt);
            \fill (4.70,0.775) circle (1.5pt);
            \node[labS,anchor=south,yshift=2pt] at (5.10,2.10) {$c_i$};
            \node[labS,anchor=north,yshift=-5pt] at (4.70,0.70) {$c_j$};

            \draw[decorate,decoration={brace,amplitude=4.5pt}] (1.0,3.30) -- (9.2,3.30);
            \node[labS,anchor=south,yshift=2pt] at (5.10,3.30) {$w_i$};

            \draw[decorate,decoration={brace,mirror,amplitude=4.5pt}] (3.4,-0.65) -- (6.0,-0.65);
            \node[labS,anchor=north,yshift=-2pt] at (4.70,-0.65) {$w_j$};
        \end{tikzpicture}
        \caption{\capj{{Complete inclusion:} $S_j\subset S_i$ so $S_{ij}=S_j$ and $\ell_{ij}=w_j$.}}
    \end{subfigure}

    \vspace{1.2em}

    \begin{subfigure}[t]{0.49\textwidth}
        \centering
        \begin{tikzpicture}[x=0.78cm,y=0.98cm,>=Latex, every node/.style={font=\footnotesize}]
            \draw[->,line width=0.8pt] (0,0) -- (10.35,0) node[right] {$r$};
            \node[labS,anchor=north,yshift=-2pt] at (0,0) {$0$};

            \fill[stateFill] (1.0,1.70) rectangle (3.8,2.35);
            \fill[stateFill] (6.2,0.45) rectangle (9.0,1.10);
            \draw[stateLine] (1.0,1.70) rectangle (3.8,2.35);
            \draw[stateLine] (6.2,0.45) rectangle (9.0,1.10);

            \node[lab] at (1.55,2.03) {$S_i$};
            \node[lab] at (8.55,0.78) {$S_j$};

            \node[lab,align=center] at (5.00,1.32) {$S_{ij}=\varnothing$\\[2pt]$\ell_{ij}=0$};

            \fill (2.40,2.025) circle (1.5pt);
            \fill (7.60,0.775) circle (1.5pt);
            \node[labS,anchor=south,yshift=2pt] at (2.40,2.10) {$c_i$};
            \node[labS,anchor=north,yshift=-2pt] at (7.60,0.70) {$c_j$};

            \draw[decorate,decoration={brace,amplitude=4.5pt}] (1.0,3.30) -- (3.8,3.30);
            \node[labS,anchor=south,yshift=2pt] at (2.40,3.30) {$w_i$};

            \draw[decorate,decoration={brace,mirror,amplitude=4.5pt}] (6.2,-0.65) -- (9.0,-0.65);
            \node[labS,anchor=north,yshift=-2pt] at (7.60,-0.65) {$w_j$};
        \end{tikzpicture}
        \caption{\capj{No overlap: $S_i\cap S_j=\varnothing$ implies $\ell_{ij}=0$ and strict selection rules.}}
    \end{subfigure}
    \hfill
    \begin{subfigure}[t]{0.49\textwidth}
        \centering
        \begin{tikzpicture}[x=0.78cm,y=0.98cm,>=Latex, every node/.style={font=\footnotesize}]
            \draw[->,line width=0.8pt] (0,2.40) -- (10.35,2.40) node[right] {$r_1$};
            \draw[->,line width=0.8pt] (0,0.00) -- (10.35,0.00) node[right] {$r_2$};

            \fill[stateFill] (1.0,2.85) rectangle (7.2,3.45); 
            \fill[stateFill] (3.2,2.20) rectangle (9.2,2.80); 
            \fill[overlapFill] (3.2,2.85) rectangle (7.2,3.45);
            \fill[overlapFill] (3.2,2.20) rectangle (7.2,2.80);
            \draw[stateLine] (1.0,2.85) rectangle (7.2,3.45);
            \draw[stateLine] (3.2,2.20) rectangle (9.2,2.80);
            \node[labS,anchor=east] at (0.95,3.15) {$i$};
            \node[labS,anchor=east] at (3.15,2.50) {$k$};
            \node[labS] at (5.20,3.62) {$S_{ik}$};

            \fill[stateFill] (2.2,0.55) rectangle (6.3,1.15); 
            \fill[stateFill] (4.4,-0.10) rectangle (10.0,0.50); 
            \fill[overlapFill] (4.4,0.55) rectangle (6.3,1.15);
            \fill[overlapFill] (4.4,-0.10) rectangle (6.3,0.50);
            \draw[stateLine] (2.2,0.55) rectangle (6.3,1.15);
            \draw[stateLine] (4.4,-0.10) rectangle (10.0,0.50);
            \node[labS,anchor=east] at (2.15,0.85) {$j$};
            \node[labS,anchor=east] at (4.35,0.20) {$l$};
            \node[labS] at (5.35,-0.35) {$S_{jl}$};

            \fill (5.20,2.55) circle (1.5pt);
            \fill (5.55,0.75) circle (1.5pt);
            \node[labS,anchor=south west,xshift=10pt,yshift=4pt] at (5.24,2.58) {$r_1\in S_{ik}$};
            \node[labS,anchor=north west,xshift=10pt,yshift=-4pt] at (5.59,0.72) {$r_2\in S_{jl}$};

            \draw[-{Latex[length=3mm]},line width=0.8pt] (5.20,2.50) -- (5.55,0.84);
            \node[labS,anchor=west,xshift=8pt] at (5.74,1.70) {$V(r_2-r_1)$};

        \end{tikzpicture}
        \caption{\capj{Two overlap intervals in $\bar J_{ij}^{kl}$: $\psi_i^*(r_1)\psi_k(r_1)\neq0$ only on $S_{ik}$ and $\psi_j^*(r_2)\psi_l(r_2)\neq0$ only on $S_{jl}$. Hence $\bar J_{ij}^{kl}=0$ if $S_{ik}=\varnothing$ or $S_{jl}=\varnothing$.}}
    \end{subfigure}

    \caption{\capj{Rectangular localized states as intervals $S_i=[c_i-w_i/2,\,c_i+w_i/2]$ and their intersections $S_{ij}=S_i\cap S_j=[S_{ij}^-,S_{ij}^+]$ with overlap length $\ell_{ij}=|S_{ij}|$. Panels (a)-(c) show partial overlap, full inclusion, and no overlap (empty intersection). The overlap interval is shaded. Panel (d) illustrates how a four-fermion matrix element samples the interaction kernel over\,two\,overlap intervals via $\mathcal{I}_V(S_{ik},S_{jl})$.}}
    \label{fig:localized_overlap}
\end{figure*}
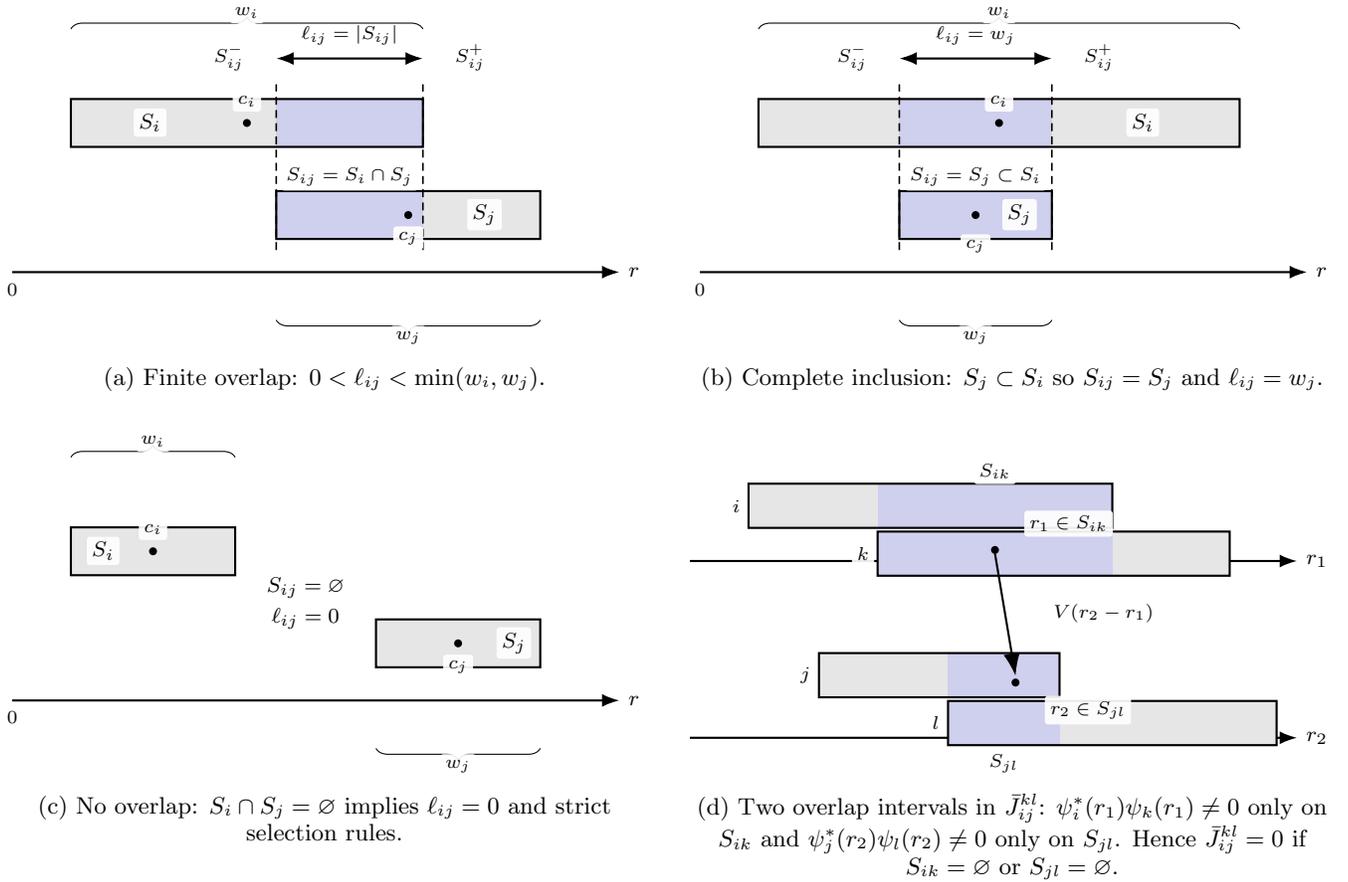

To this end, we define the real-space overlap interval $S_{ij}=S_i\cap S_j$, which will be used throughout (see Fig.~\ref{fig:localized_overlap} ). To avoid confusion, we denote the Hilbert-space overlap matrix by
\begin{equation}
\Omega_{ij}\equiv\int_0^Ldr\,\psi_i^*(r)\psi_j(r)
=\delta_{ij}+(1-\delta_{ij})\,\frac{\ell_{ij}}{\sqrt{w_iw_j}}\,e^{i(\varphi_j-\varphi_i)},
\label{eq:seed_overlap_matrix}
\end{equation}
where $\ell_{ij}\equiv |S_i\cap S_j|$. The canonical orbitals entering Eq.~(\ref{eq:syk}) are obtained by symmetric orthogonalization,
\begin{equation}
\phi_i(r)=\sum_j (\Omega^{-1/2})_{ji}\,\psi_j(r),
\qquad
\int_0^Ldr\,\phi_i^*(r)\phi_j(r)=\delta_{ij}.
\label{eq:lowdin_orbitals}
\end{equation}
\begin{equation}
\bar J_{ij}^{kl,\mathrm{can}}=
\int dr_1dr_2\,\phi_i^*(r_1)\phi_j^*(r_2)\phi_k(r_1)\phi_l(r_2)\,V(r_2-r_1).
\label{eq:J_can_exact}
\end{equation}
The seed envelopes and the canonical orbitals play different roles. The functions $\psi_i$ are auxiliary localized envelopes used to expose the overlap geometry. If one defines projected operators $d_i=\int dr\,\psi_i^*(r)\hat\Psi(r)$ from the microscopic fermion field $\hat\Psi(r)$, then $\{d_i,d_j^\dagger\}=\Omega_{ij}$, so these operators are not canonical when the envelopes overlap. Canonical fermions are instead built from the L\"owdin-orthogonalized orbitals, $c_i=\int dr\,\phi_i^*(r)\hat\Psi(r)$, for which $\{c_i,c_j^\dagger\}=\delta_{ij}$. Moreover, the spatial overlap is allowed. The requirement is orthonormality, not disjoint support. Equation~(\ref{eq:J_can_exact}) is therefore the exact quartic matrix element in the canonical basis. For the geometric discussion below we also use the auxiliary seed tensor $\bar J_{ij}^{kl}$ built directly from the nonorthogonal envelopes. The canonical tensor is obtained from it by four factors of $\Omega^{-1/2}$. Writing $\Omega=I+\delta\Omega$ gives $\Omega^{-1/2}=I-\tfrac12\delta\Omega+O(\delta\Omega^2)$, so orthogonalization changes a typical matrix element through the off-diagonal overlaps. At the single-patch level these corrections need not be small, so this section should be read as the geometric baseline. The controlled canonical limit is reached later, when internal phase randomization makes $\Omega-I$ perturbatively small.
\begin{figure*}
    \centering
    \includegraphics[width=.4\linewidth]{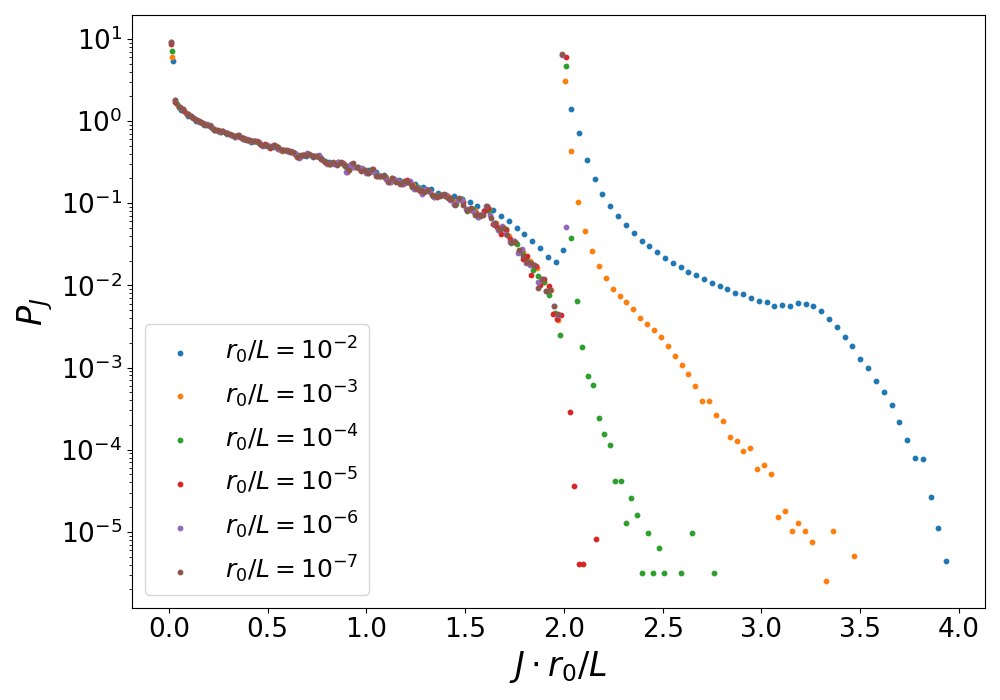}
    \hspace{.05\linewidth}
    \includegraphics[width=.4\linewidth]{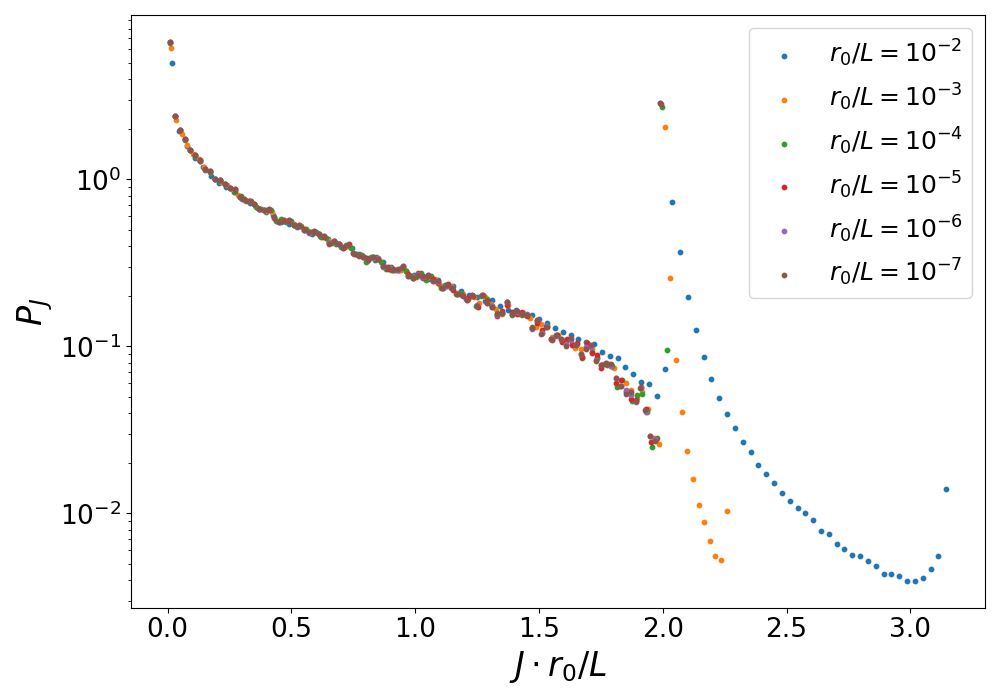}
    \caption{\capj{Representative histograms of the nonzero values of $\Re J_{ij}^{kl}$ and $\Im J_{ij}^{kl}$ for the variable-width ensemble $w_i\in[0,\overline{w}L]$ (left) and the constant-width ensemble $w_i=\overline{w}L$ (right), shown for several values of the short-distance cutoff $r_0$. Once $r_0\ll \overline{w}L$, the curves nearly collapse, indicating that the statistics of the induced couplings are controlled primarily by overlap geometry rather than by the precise ultraviolet regularization. Yje figure displays one Cartesian projection of a rotationally invariant ensemble whose underlying radial distribution is the gauge-invariant object.}}
    \label{fig:J_hist}
\end{figure*}

For the geometric analysis, it is convenient to begin with the seed-basis tensor
\begin{equation}
    \bar{J}_{ij}^{kl} = \int dr_1 dr_2
    \psi_i^*(r_1) \psi_j^*(r_2) \psi_k(r_1) \psi_l(r_2) V(\Delta r)
    \label{eq:Jbar_def}
\end{equation}
constructed directly from the seed envelopes, i.e., the nonorthogonal states. Eq.~(\ref{eq:Jbar_def}) has the same algebraic form as Eq.~(\ref{eq:J_can_exact}), but a different status: Eq.~(\ref{eq:J_can_exact}) is the exact canonical matrix element, whereas Eq.~(\ref{eq:Jbar_def}) is an auxiliary object that makes the real-space overlap geometry explicit. Here and below the integrals run over $r_1,r_2\in[0,L]$, with $\Delta r=r_2-r_1$. Since we work at $\overline{w}\ll1$, one may equally well extend them to $(-\infty,\infty)$, understanding that the orbitals vanish outside the sample. It is immediate that $\bar J_{ij}^{kl}=\bar J_{kl}^{ij*}$. To keep the notation light, we denote the seed tensor simply by $\bar J_{ij}^{kl}$ throughout the geometric analysis, and we restore the superscript `can' only when the exact orthogonalized tensor must be displayed explicitly. We will mainly focus on the regularized Coulomb potential $V_c$,
but the local uniform potential $V_u(r)$ will also be considered.
\begin{equation}
\begin{split}
V_c(r) &= 1 / \max(|r|, r_0) \squad,\\ 
V_u(r) &= {\theta(r_0-|r|)\equiv \theta(-\bar{r})\squad,\quad \bar{r}\equiv |r|-r_0}
\end{split}
\end{equation}
The parameter $r_0$ plays the role of a microscopic short-distance cutoff: for $V_c$, it regularizes the $1/|r|$ singularity (e.g., due to a finite transverse width, lattice spacing, or screening at short distances), while for $V_u$ it sets an explicit interaction range (a coarse-grained model of screened interactions). In the numerical simulations, we will focus on the regime $r_0\ll \overline{w}L$ where the statistics of the induced couplings become insensitive to the precise cutoff.

As the eigenstates \refe{eigens} are piecewise constant on their support,
it is convenient to encode their real-space structure by intervals.
Define the support of orbital $i$ as $S_i\equiv[c_i-w_i/2,\,c_i+w_i/2]$,
and the pairwise overlap interval as $S_{ab}\equiv S_a\cap S_b=[S_{ab}^-,S_{ab}^+]$
with $S_{ab}^-\equiv\max(c_a-w_a/2,\,c_b-w_b/2)$ and 
$S_{ab}^+\equiv\min(c_a+w_a/2,\,c_b+w_b/2)$.
$S_{ab}=\varnothing$ when $S_{ab}^- \geq S_{ab}^+$.

The product $\psi_i^*(r)\psi_k(r)$ is the constant $e^{i(\varphi_k-\varphi_i)}/\sqrt{w_i w_k}$
within $S_{ik}$ (and is zero outside of it),
and similarly $\psi_j^*(r)\psi_l(r)=e^{i(\varphi_l-\varphi_j)}/\sqrt{w_j w_l}$ for $S_{jl}$. Therefore, the matrix element factorizes into a universal phase and an interval--interval interaction energy. We make this explicit by introducing the functional
\begin{equation}
\label{eq:JV_def}
\mathcal{I}_V(S_1,S_2)\equiv \int_{S_1} dr_1\int_{S_2} dr_2 V(\Delta r),
\end{equation}
defined for any two (possibly empty) intervals $S_1,S_2\subset(-\infty,\infty)$. With this notation, \refe{eigens} yields the compact form
\begin{equation}
\label{eq:Jbar_compact}
\bar J_{ij}^{kl}=\frac{e^{i\Phi_{ijkl}}}{\sqrt{w_i w_j w_k w_l}} \mathcal{I}_V(S_{ik},S_{jl})
\end{equation}
with $\Phi_{ijkl}\equiv \varphi_k+\varphi_l-\varphi_i-\varphi_j$.
This immediately shows that $\bar J_{ij}^{kl}=0$ whenever either $S_{ik}$ or $S_{jl}$ is empty. The real part in Eq. (\ref{eq:Jbar_compact}) is determined only by the geometric disorder. Under a change of orbital phases,
$c_i \rightarrow e^{i\chi_i} c_i,$
the interaction matrix element transforms as
$J_{ij}^{kl}\rightarrow e^{i(\chi_k+\chi_l-\chi_i-\chi_j)}J_{ij}^{kl}.$
Therefore, for a fixed realization, the separate quantities $\Re J_{ij}^{kl}$ and $\Im J_{ij}^{kl}$ are not gauge invariant. In the present single-patch form, however, the phase $\Phi_{ijkl}$ is only an overall rephasing of the orbitals: by choosing $\chi_i=-\phi_i$, one can remove all these phases at once, so that the seed couplings become real. Thus, at this level, the physically meaningful information is carried by the amplitudes $ |J_{ij}^{kl}|$ and by the pattern of exact zeros.

This changes once each localized state is resolved into many pieces with different internal random phases. In that case, each $J_{ij}^{kl}$ becomes a sum of many contributions with different relative phases. As long as these internal phases cannot be reduced to a single overall phase for each state, one phase choice $\chi_i$ per orbital is no longer enough to remove them all. The resulting phase structure is then genuine $U(1)$ disorder and cannot, in general, be gauged away.

A useful estimate already appears at the level of interval overlaps. For two independent intervals of typical width $\overline{w}L$ with uniformly random centers, the probability of a nonempty overlap is $1-(1-\overline{w})^2\simeq 2\overline{w}$ for $\overline{w}\ll 1$. Since $\bar J_{ij}^{kl}$ is nonzero only when both overlap intervals $S_{ik}$ and $S_{jl}$ are nonempty, a typical induced coupling has a finite probability of vanishing exactly. This microscopic geometric sparsity is the origin of disconnected components (SYK clusters) in the interaction graph at finite $N$, and it underlies the percolation-like saturation behavior observed as $N$ increases.

From the viewpoint of the induced SYK disorder, \refe{Jbar_compact} separates the two sources of randomness very clearly.  
All dependence on the random phases $\{\varphi_\alpha\}$ enters only through $e^{i\Phi_{ijkl}}$, which is a pure U(1) phase.  
All dependence on the geometric disorder $\{(c_\alpha,w_\alpha)\}$ enters only through the real functional $\mathcal{I}_V(S_{ik},S_{jl})$ and the prefactor $1/\sqrt{w_i w_j w_k w_l}$.  
This factorization is the microscopic origin of the rotational invariance of the coupling distribution discussed in the next \refs{distJ}.

For later use it is helpful to evaluate $\mathcal{I}_V(S_1,S_2)$ in closed form. Introduce a second primitive $I(r)$ of $V(r)$, defined by
\begin{equation}
   I''(r)=V(r) \squad.
\end{equation}
The freedom $I(r)\rightarrow I(r)+ar+b$ does not affect any physical quantity below, because $\mathcal{I}_V$ only depends on finite differences of $I$ at the interval endpoints. 
Performing the $r_2$ integral in \refe{JV_def} first gives
\begin{equation*}
\int_{S_2} dr_2 V(r_2-r_1) =I'(S_2^+-r_1)-I'(S_2^--r_1)\squad,
\end{equation*}
and hence the exact boundary-term expression of $\mathcal{I}_V$ becomes
\begin{align}
\label{eq:JV_endpoints}
\mathcal{I}_V(S_1,S_2)=&\int_{S_1} dr_1 \Big[I'(S_2^+-r_1)-I'(S_2^- - r_1)\Big]=\nonumber\\
&\Big[-I(S_2^+-r_1)+I(S_2^- - r_1)\Big]_{r_1=S_1^-}^{r_1=S_1^+}=\\
I(S_2^+ - S_1^-) + &I(S_2^- - S_1^+) - I(S_2^+ - S_1^+) - I(S_2^- - S_1^-)\ .\nonumber
\end{align}
The obtained expression does not depend on the ambiguity $I(r)\rightarrow I(r)+ar+b$
as \refe{JV_endpoints} contains only endpoint differences of $I$.  



For $I_c$ and $I_u$ which correspond to $V_c$ and $V_u$, respectively, we get
\begin{equation}
\begin{split}
    I_c(r) &=
    \theta(-\bar{r})\frac{r^2}{2 r_0} + \theta(\bar{r})\cdot \inb({|r| \ln \frac{|r|}{r_0} + \frac{r_0}{2}}) \squad,\\
    I_u(r) &=
    \theta(-\bar{r}) \frac{r^2}{2} + \theta(\bar{r}) (|r|+\bar{r}) \frac{r_0}{2} \squad.
\end{split}
\end{equation}
The additive and linear (ambiguous) pieces of $I_c$ and $I_u$
are fixed by requiring $I_{c/u}(0)=0$ and 
$I_{c/u}(r) = I_{c/u}(-r)$. The latter can be achieved for all even potentials $V(r)=V(-r)$.

The two primitives exhibit qualitatively different long-distance behavior. For the regularized Coulomb potential, one has $I_c(r)\sim |r|\ln(|r|/r_0)$ for $|r|\gg r_0$, reflecting the slow $1/|r|$ decay and yielding broad coupling amplitudes even for well-separated overlap intervals. By contrast, $I_u(r)$ is asymptotically linear, $I_u(r)\sim r_0|r|$ for $|r|\gg r_0$, consistent with a finite interaction range. This tends to increase the probability of exactly vanishing couplings and to bound the support of the amplitude distribution $P_A$ derived below.

\subsection{Probability distribution of the induced couplings}
\label{sec:distJ}

Because many distinct $J_{ij}^{kl}$ depend on the same orbital data, the full interaction tensor contains strong inter-coupling correlations. The simplest object to analyze is nevertheless the single-coupling marginal distribution. It already contains the two qualitative features that matter most: a finite point mass at $J=0$ and a broad, generally non-Gaussian regular sector.

The couplings $J_{ij}^{kl}$ in \refe{syk} are induced by the random orbital parameters in \refe{eigens}, so their statistics are determined by the microscopic geometric disorder rather than postulated independently. This is already visible in the numerical histograms of \reff{J_hist}. We now formulate the disorder average in a way that makes the phase-amplitude structure explicit. Later in Sec .~\ref {sec:true_syk}, we will contrast the resulting one-point law with the Gaussian SYK limit.

We parametrize  each single-particle state by $(c_i,w_i,\varphi_i)$, where $c_i\in[0,L]$ and $\varphi_i\in[0,2\pi)$, which are IID uniform random variables. For the widths, we consider two ensembles (already specified above): (i) a constant-width ensemble $w_i=\overline{w}L$, and (ii) a variable-width ensemble with $w_i$ IID uniform on $[0,\overline{w}L]$. It is convenient to encode both cases by width probability densities $p_w$
\begin{equation}
\begin{split}
p_w(w)=\delta(w-\overline{w}L) \quad &\text{for } w_i=\overline{w}L \squad,\\
p_w(w)=\frac{1}{\overline{w}L}\,\theta(w)\,\theta(\overline{w}L-w) 
\quad &\text{for } w_i\in[0,\overline{w}L]\squad.
\end{split}
\label{eq:pw_def}
\end{equation}
Then, for any $n$-index observable $\mathcal{O}$
defined on a set of indices $\{i_1,\dots,i_n\}$,
one can write the disorder average as
\begin{equation}
\label{eq:disorder_measure}
\inb<{\mathcal{O}}> \equiv \prod_{q=1}^n
\inb[{
\int\limits_0^{L}\frac{dc_{i_q}}{L}
\int\limits_0^{2\pi}\frac{d\varphi_{i_q}}{2\pi}
\int\limits_0^{\infty}dw_{i_q} p_w(w_{i_q})
}]\mathcal{O}\squad.
\end{equation}
For distinct indices $(i,j,k,l)$, we define the disorder-averaged probability density of the (generally complex) coupling as
\begin{equation}
\label{eq:PJ_def}
P_J(J)\equiv 
\Big\langle 
\delta\!\left(\Re J-\Re J_{ij}^{kl}\right)\,
\delta\!\left(\Im J-\Im J_{ij}^{kl}\right)
\Big\rangle \squad,
\end{equation}
where $J_{ij}^{kl}$ is the antisymmetrized coefficient in \refe{syk} and $\Re\ldots$ and $\Im\ldots$ stand for the real and imaginary parts. \refe{PJ_def} defines a probability density on the complex plane. For any sufficiently regular test function $f$, one has
\begin{align*}
\inb<{f(J_{ij}^{kl})}>
&=\int d^2 J P_J(J) f(J)\squad,\\
\int d^2 J\,P_J(J)&=1\quad,\quad d^2J\equiv d(\Re J)\,d(\Im J)\squad.
\end{align*}
Because the orbital parameters are IID and exchangeable, the single-coupling distribution $P_J$ is the same for any choice of distinct indices $(i,j,k,l)$. In practice, numerical histograms of $J_{ij}^{kl}$ represent Monte Carlo estimates of $P_J$ obtained by sampling many index quadruples and disorder realizations. It is also convenient to introduce the compact notation $\delta^{(2)}(J)\equiv\delta(\Re J)\,\delta(\Im J)$, so that \refe{PJ_def} can be rewritten as $P_J(J)=\langle \delta^{(2)}(J-J_{ij}^{kl})\rangle$. This emphasizes that $P_J$ is the pushforward of the microscopic disorder measure \refe{disorder_measure} under the map $(c,w,\varphi)\mapsto J_{ij}^{kl}$. 

We now use \refee{JV_def}{Jbar_compact} to express the antisymmetrized couplings in a phase--amplitude form and to derive their probability distributions.

For the two interaction potentials studied here, $V(r)=V(-r)$ is real and even. Therefore, $\mathcal{I}_V(S_1,S_2)=\mathcal{I}_V(S_2,S_1)$. 
The symmetry follows from swapping $r_1\leftrightarrow r_2$ in \refe{JV_def} and
using $V(r_2-r_1)=V(r_1-r_2)$.
The same change of variables implies $\bar J_{ij}^{kl}=\bar J_{ji}^{lk}$,
so the full antisymmetrization
$J_{ij}^{kl}=\bar J_{ij}^{kl}-\bar J_{ji}^{kl}-\bar J_{ij}^{lk}+\bar J_{ji}^{lk}$
reduces to $2(\bar J_{ij}^{kl}-\bar J_{ji}^{kl})$, and thereby
\begin{equation}
\begin{split}
\label{eq:J_phase_amp}
J_{ij}^{kl}
&=e^{i\Phi_{ijkl}}A_{ij}^{kl} \squad,\\
A_{ij}^{kl} &= 2 \cdot \frac{\mathcal{I}_V(S_{ik},S_{jl})-\mathcal{I}_V(S_{jk},S_{il})}{\sqrt{w_i w_j w_k w_l}}
\end{split}
\end{equation}
where the amplitude $A_{ij}^{kl}$ is real and depends only on the geometric disorder $(c_\alpha,w_\alpha)$.
Note that $A_{ij}^{kl}$ can have either sign because it is a difference of two interval--interval interaction energies.  
However, since $\Phi_{ijkl}$ is uniform, the distribution of the complex product $e^{i\Phi_{ijkl}}A_{ij}^{kl}$ depends only on the radius $|A_{ij}^{kl}|$; any sign of $A_{ij}^{kl}$ can be absorbed into a shift $\Phi_{ijkl}\rightarrow \Phi_{ijkl}+\pi$.

Since the phases $\varphi_\alpha$ are IID and uniform, the combination $\Phi_{ijkl}$ is uniform on $[0,2\pi)$ and independent of $A_{ij}^{kl}$. 
Hence $P_J(J)$ is rotationally invariant in the complex plane, i.e. $P_J(J)=P_J(|J|)$.
It can be split into two terms:
(i) a singular term describing the probability $P_0$ of an exactly vanishing coupling and
(ii) a regular term of the probability density $p_J(J)$ when $|J|>0$.
$P(J)$ can be further mapped to the probability density $P_A(A)$
of the nonnegative amplitude $A=|J|$.
The latter plays the role of an ``effective interaction-scale distribution'' generated by random overlaps and the interaction kernel, while $P_0$ quantifies geometric sparsity and controls percolation in the interaction graph.
The discussed distributions are thus
\begin{equation}
\begin{split}
P_J(J) = &P_0 \delta^{(2)}(J) + p_J(J) \squad,\\
P_A(A) = &P_0 \delta^{(1)}(A) + p_A(A)
\end{split}
\label{eq:PJA}
\end{equation}
and their regular sectors are connected through simple reparametrization as
\begin{equation}
p_J(J)=\frac{1}{2\pi|J|} p_A(|J|) \squad.
\end{equation}
The first term with $P_0=\mathrm{Prob}(|J|=0)$ yields the ``point mass" at the origin. 
The assumed normalizations are $\int_0^{\infty}dA\,p_A(A)=\int\int dJ p_J(J)=1-P_0$. The point mass at $J=0\ (A=0)$ is explicitly accounted for by $P_0$.
Physically, $P_0$ quantifies geometric sparsity: a finite fraction of couplings vanish identically because the required overlap intervals are empty (and, for short-range interactions, because the overlap intervals can lie outside the interaction range).

For later use, we define the \emph{active sector} distribution by conditioning on nonzero couplings,
\begin{equation}
P_J^{\rm act}(J)\equiv \frac{p_J(J)}{1-P_0},
\qquad
\int d^2J\,P_J^{\rm act}(J)=1.
\end{equation}
Equivalently, the active sector is the full distribution with the point mass at $J=0$ removed and the remainder renormalized. In a given disorder realization, the corresponding set of nonzero couplings will be called the active set.


To connect $P_J(J)$ to the standard one-dimensional histograms of $\Re J$ and $\Im J$, we integrate the two-dimensional density over the orthogonal direction.
For $x\equiv \Re J$ (and equivalently for $\Im J$) this marginal is
\begin{equation*}
p_{\Re J}(x)=\int_{-\infty}^{\infty}dy\,p_J(x+iy) \squad.
\end{equation*}
The point mass in \refe{PJA} contributes an additional term $P_0\,\delta(x)$,
$P_{\Re J}(x)=P_0\,\delta(x)+p_{\Re J}(x)$.
The connection to $P_A$ is given through
\begin{equation}
\label{eq:pReJ_Abel}
P_{\Re J}(x)=P_0 \delta(x)
+\int_{|x|}^{\infty}\frac{dA}{\pi\sqrt{A^2-x^2}}\,p_A(A),
\end{equation}
and the same expression holds for $\Im J$ by symmetry.

\refe{pReJ_Abel} is the standard Abel transform relating an isotropic two-dimensional distribution to its Cartesian marginal.  
In particular, the square-root kernel implies a characteristic nonanalyticity in $p_{\Re J}(x)$ at the edge of the support of $p_A(a)$. Conversely, one may reconstruct $p_A$ from $p_{\Re J}$ by the inverse Abel transform.


\subsection{Consequences of width disorder}
\label{sec:width_disorder}

The random width ensemble enters only through the measure \refe{disorder_measure}. 
For later reference, we record the overlap-length statistics in the constant-width case, which provides a useful analytic benchmark for numerical sampling. 

The statistics of the induced couplings are controlled by the geometry of interval overlaps.  
A basic building block is the (nonnegative) overlap length between two orbitals,
\begin{align*}
S_{ij}&\equiv S_i\cap S_j=[S_{ij}^-,S_{ij}^+]\squad,\\
\ell_{ij}&\equiv |S_{ij}|=\max\inb({0,S_{ij}^+ - S_{ij}^-})
\end{align*}
with $S_i$ and the overlap endpoints $S_{ij}^\pm$ defined in \refs{geometric_construction}.  
Physically, $\ell_{ij}$ is the real-space phase space on which the bilinear $\psi_i^*(r)\psi_j(r)$ is nonzero: for the rectangular orbitals in \refe{eigens} this bilinear is constant on $S_{ij}$ and vanishes identically outside.  
Thus $\ell_{ij}=0$ is a strict selection rule: if two orbitals do not overlap, any matrix element that requires their overlap vanishes exactly.  
In particular, for long-range interactions such as $V_c$, a nonzero unsymmetrized coupling $\bar J_{ij}^{kl}$ requires both overlap intervals $S_{ik}$ and $S_{jl}$ in \refe{Jbar_compact} to be nonempty.  
The overlap-length distribution, therefore, directly quantifies the geometric sparsity that produces disconnected components in the interaction graph, and it provides a minimal analytic handle on how width disorder modifies the prevalence of nonzero induced couplings.

Mathematically, for two independently sampled orbitals $i$ and $j$ drawn from the disorder measure \refe{disorder_measure}, $\ell_{ij}$ is a random variable with a \emph{mixed} distribution: there is generally a finite probability mass at $\ell=0$ and a continuous density for $\ell>0$.  
We define $p_\ell(\ell)$ to be the regular part, i.e.
\begin{equation*}
p_\ell(\ell)\equiv \Big< \delta\inb({\ell-\ell_{ij}}) \Big>_{c,w}
\quad,\quad \ell>0 \squad,
\end{equation*}
while the atom at the origin is recorded separately as $\pi_0=\mathrm{Prob}(\ell_{ij}=0)$.  
Equivalently, the full probability measure for $\ell_{ij}$ can be written as
\begin{equation*}
P_\ell(\ell)=\pi_0 \delta(\ell)+p_\ell(\ell) 
\end{equation*}
with normalization
\begin{equation*}
\int_{0}^{\infty}\!d\ell\,p_\ell(\ell)=1-\pi_0\squad.
\end{equation*}


\subsubsection{Constant-width benchmark ($w_i=\overline{w}L$).}

For two independent centers $c_1,c_2$ uniform on $[0,L]$, the absolute separation 
$d\equiv|c_1-c_2|$ has the triangular density $p_d(d)=2(L-d)/L^2$ on $0\le d\le L$.  
For given widths $w_1$ and $w_2$
the overlap length is given by
\begin{equation}
\ell=\left\{
\begin{tabular}{l@{\squad,\squad}c@{ $2d$ }c}
    $0$              & $\Sigma w <$ & \\
    $\Sigma w/2 - d$ & $\Delta w <$ & $< \Sigma w$ \\
    $w_<$            &              & $< \Delta w$
\end{tabular}\right.
\label{eq:overlap_cw}
\end{equation}
with the notations $\Sigma w \equiv w_1+w_2$, $\Delta w \equiv |w_1-w_2|$,
$w_<=\min(w_1,w_2)$.
Using these relations alongside with the distribution $p_d$ of $d$
one can derive the distribution $P_\ell$ to be
\begin{equation}
\begin{split}
P_{\bar\ell}(\bar\ell|w_1,w_2)= &\delta(\bar\ell) \cdot (1-\Sigma w/2L)^2 +\\
\theta(w_</L-\bar\ell) &\theta(\bar\ell) \cdot \inb({2-\frac{\Sigma w}{L}+2\bar\ell})+\\
\delta(w_</L-\bar\ell) &\cdot \frac{\Delta w}{2L} \inb({2 - \frac{\Delta w}{2L}})
\end{split}
\label{eq:overlap_distr_cw}
\end{equation}
where we have introduced the notation $\overline{\ell}=\ell/L$
and the corresponding distribution $P_{\bar\ell}(\bar\ell)=LP_\ell(l)$.
In the case $w_i=w_j=\overline{w}L$ the expression reduces to
\begin{equation}
\pi_0=(1-\overline{w})^2\quad,\quad
p_{\bar\ell}({\bar\ell})=\theta(\overline{w}-\bar\ell) \cdot 2(1-\overline w + \bar\ell)
\label{eq:overlap_distr_ecw}
\end{equation}
as in that case $\Delta w=0$ and $w_</L=\overline{w}$.


\subsubsection{Variable-width ensemble.}
For the variable-width ensemble with IID widths $w_i\sim{\rm Unif}[0,\overline{w}L]$, the overlap-length distribution can be obtained by averaging the conditional law \refe{overlap_distr_cw} over $w_1,w_2$. Below, we instead give a direct geometric derivation that separates partial-overlap and full-inclusion configurations and yields the closed forms.

In the variable-width ensemble, the overlap length is no longer a simple linear function of $d$ because one interval can be fully contained inside the other.  
Introduce the ordered widths $w_{>}\equiv\max(w_i,w_j)$ and $w_{<}\equiv\min(w_i,w_j)$.  
For a fixed separation $d=|c_i-c_j|$ the overlap length is
\begin{eqnarray*}
&&\ell(d;w_{>},w_{<})=\\
&&=\begin{cases}
0, & d\ge \dfrac{w_{>}+w_{<}}{2},\\[0.4em]
w_{<}, & d\le \dfrac{w_{>}-w_{<}}{2},\\[0.6em]
\dfrac{w_{>}+w_{<}}{2}-d, & \dfrac{w_{>}-w_{<}}{2}< d < \dfrac{w_{>}+w_{<}}{2}.
\end{cases}
\end{eqnarray*}

The point mass at $\ell=0$ is controlled only by the condition for a nonempty intersection, $d<(w_i+w_j)/2$.  
Using the triangular $p_d(d)$ one finds
\begin{eqnarray*}
\mathrm{Prob}(\ell=0\,|\,w_i,w_j)&=&\int_{(w_i+w_j)/2}^{L}\!dd\,p_d(d)\\
&=&\left(1-\frac{w_i+w_j}{2L}\right)^2,
\end{eqnarray*}
and averaging over two IID widths $w_i,w_j$ uniform on $[0,W]$ yields the closed form for $\mathrm{Prob}(\ell_{ij}=0)$ and $p_\ell(\ell)$.  
For $\ell>0$ one must include both partial overlaps and the full-inclusion regime in the calculation of $p_\ell(\ell)$.  
A convenient way to organize the calculation is to work with ordered widths $(w_{>},w_{<})$ whose joint density is $2/W^2$ on $0<w_{<}<w_{>}<W$.  
The continuous density can then be written as a sum of two elementary contributions,
\begin{align*}
p_\ell^{(\mathrm{partial})}(\ell)
&=\frac{2}{W^2}\int_{\ell}^{W}\!dw_{<}\int_{w_{<}}^{W}\!dw_{>}\;
p_d\!\left(\frac{w_{>}+w_{<}}{2}-\ell\right),\\[0.4em]
p_\ell^{(\mathrm{incl})}(\ell)
&=\frac{2}{W^2}\int_{\ell}^{W}\!dw_{>}\;
\int_{0}^{(w_{>}-\ell)/2}\!dd\;p_d(d),
\end{align*}
so that $p_\ell(\ell)=p_\ell^{(\mathrm{partial})}(\ell)+p_\ell^{(\mathrm{incl})}(\ell)$.  
Carrying out the remaining polynomial integrals yields the compact factorized expression:

\begin{equation}
\label{eq:overlap_P0_vw}
\mathrm{Prob}(\ell_{ij}=0)=1-\overline{w}+\frac{7}{24}\overline{w}^2,
\end{equation}
\begin{eqnarray}
\label{eq:overlap_pdf_vw}
p_\ell(\ell)\!&=&\!\frac{(W-\ell)^2\left(18L-7W+7\ell\right)}{6L^2W^2}\\
&\!=\!&\frac{1}{6L} \frac{(\overline{w}-\bar\ell)^2(18-7\overline{w}+7\bar\ell)}{\overline{w}^2},
\qquad 0<\ell<W,\nonumber
\end{eqnarray}
where $W\equiv\overline{w}L$.

Two remarks clarify the physical role of width disorder.  
First, in the dilute limit $\overline{w}\ll 1$, \refe{overlap_P0_vw} gives $\mathrm{Prob}(\ell>0)=1-\mathrm{Prob}(\ell=0)\simeq \overline{w}$, parametrically smaller than the constant-width result $2\overline{w}$ in \refe{overlap_distr_ecw}.  Intuitively this reflects that the typical width in the uniform $[0,W]$ ensemble is only $W/2$, so typical pairs overlap less often.  
Second, \refe{overlap_pdf_vw} contains the factor $(W-\ell)^2$, showing that near-perfect overlaps $\ell\approx W$ are strongly suppressed: producing a large intersection requires both widths to be close to $W$ and the centers to be exceptionally close.  

These overlap statistics feed directly into the coupling distribution in \refs{distJ}.  
The point mass $P_0$ in \refe{PJA} is enhanced by width disorder through the increased probability that at least one of the overlap intervals entering \refe{J_phase_amp} is empty.  
Moreover, the prefactor $1/\sqrt{w_iw_jw_kw_l}$ in \refe{J_phase_amp} implies that rare narrow orbitals can generate parametrically large couplings even when overlaps are modest (for $w_i=w_j=w_k=w_l=w$ this factor scales as $w^{-2}$), leading to broad non-Gaussian tails in $P_A(a)$ and hence in $p_{\Re J}(x)$.

\subsection{Regularized Coulomb interaction $V_c$}
\label{sec:distJ_coulomb}

For $V_c(r)=1/\max(|r|,r_0)$ the functional $\mathcal{I}_{V_c}$ in \refe{JV_def} is obtained by inserting the explicit second primitive $I_c(r)$ given above into the closed form for $\bar J_{ij}^{kl}$. 
Since $V_c$ is long ranged, $\mathcal{I}_{V_c}(S_1,S_2)$ is nonzero whenever both overlap intervals $S_1$ and $S_2$ are nonempty; thus, in this case the point mass $P_0$ in \refe{PJA} is entirely due to geometric events where at least one of the overlap intervals appearing in \refe{J_phase_amp} is empty. 
The corresponding amplitude distribution is given exactly by the finite-dimensional disorder integral
\begin{eqnarray}
\label{eq:pA_coulomb}
&&P_A(A)=
\prod_{\alpha\in\{i,j,k,l\}}
\left[\int\frac{dc_\alpha}{L}\int_0^{\infty}dw_\alpha\,p_w(w_\alpha)\right]\\
\nonumber\\
&&\times \delta \inb({A-\frac{2}{\sqrt{w_iw_jw_kw_l}} \Big|
\mathcal{I}_{V_c}(S_{ik},S_{jl})-\mathcal{I}_{V_c}(S_{jk},S_{il})
\Big| }),\nonumber
\end{eqnarray}
which, together with \refeee{PJA}{pReJ_Abel}, yields the full complex and one-dimensional distributions without approximations.


\subsection{Local uniform interaction $V_u$}
\label{sec:distJ_local}

For the finite-range interaction $V_u(r)=\theta(r_0-|r|)$ the same general framework applies, with $\mathcal{I}_{V_u}$ obtained by substituting $I_u(r)$ in the expression for $\bar J_{ij}^{kl}$. 
In contrast to the Coulomb case, the compact support of $V_u$ implies an additional geometric mechanism for exact zeros: even when both overlap intervals $S_1$ and $S_2$ are nonempty, one can have $\mathcal{I}_{V_u}(S_1,S_2)=0$ if the separation between the two intervals exceeds $r_0$. 
As a result, the weight $P_0$ in \refe{PJA} is enhanced for $V_u$, and the continuous part of $P_A(A)$ has compact support set by the maximal possible overlap of intervals within range $r_0$. 
The exact amplitude distribution is still given by the disorder integral \refe{pA_coulomb} with the replacement $\mathcal{I}_{V_c}\to\mathcal{I}_{V_u}$.

In summary, the geometric construction produces a correlated disorder ensemble: the couplings are rotationally invariant, a finite fraction are exactly zero, and the continuous nonzero part is broad and non-Gaussian. We call this continuous nonzero part the \emph{active sector}, meaning the full distribution after the point mass at $J=0$ is removed and the remainder is renormalized. In the next section we ask when this active sector becomes the Gaussian SYK law.


\section{Gaussianization of the active sector: approach to the SYK limit}
\label{sec:true_syk}

In Sec.~\ref{sec:correlated_disorder}, each orbital was represented by a single coarse support interval. That construction gives the geometric baseline: many couplings are exactly zero, the active sector is broad and non-Gaussian, and different matrix elements are correlated because they depend on the same overlap data. We now ask what extra microscopic structure is needed for this active sector to approach the canonical complex-SYK Gaussian law. This is the relevant SYK limit here, because the zero pattern fixed by real-space overlap remains present at fixed localization geometry.

In the canonical complex SYK model, the independent couplings $J_{ij}^{kl}$ ($i<j$ and $k<l$) are drawn from a zero-mean circular complex-Gaussian ensemble, subject only to hermiticity and antisymmetry,
\begin{equation}
\label{eq:gauss_target}
P_{\rm G}(J)=\frac{1}{\pi\sigma_J^2}\exp\!\left(-\frac{|J|^2}{\sigma_J^2}\right),
\qquad
\langle |J|^2\rangle=\sigma_J^2,
\end{equation}
with $J_{ij}^{kl}=-J_{ji}^{kl}=-J_{ij}^{lk}$ and $J_{ij}^{kl}=J_{kl}^{ij\ast}$; see Eq.~(\ref{eq:syk}) and Refs.~\cite{SachdevYe1993,Kitaev2015,Sachdev2015,MaldacenaStanford2016,Gu2020}. In the standard large-$N$ normalization, $\sigma_J^2=\mathcal{I}^2/N^3$, where $\mathcal{I}$ sets the interaction scale. In our construction, the statistics inside the active sector can nevertheless become canonical. Once each localization volume contains many phase-randomized microscopic pieces, a typical nonzero matrix element becomes a sum of many small complex contributions, and the active sector Gaussianizes.

The underlying mechanism is the central-limit self-averaging of complex random phases. A coupling approaches the Gaussian SYK distribution whenever it can be represented as a sum of many zero-mean contributions, $X_\alpha$, $\alpha=1\ldots M$
\begin{equation}
\label{eq:CLT_sum}
J=\sum_{\alpha=1}^{M} X_\alpha,
\qquad
\langle X_\alpha\rangle=0,
\qquad
\langle |X_\alpha|^2\rangle<\infty,
\end{equation}
with a large effective count $M\gg1$. For independent terms this is the usual two-dimensional central limit theorem. The same conclusion remains true when residual dependencies are weak enough that no small subset of terms carries a finite part of the variance. The important physical point is therefore simple: complete independence of the internal geometry is not required. What matters is that many contributions of comparable size carry independent random phases. Then the active sector becomes circular Gaussian. At the same time, connected correlations between two typical couplings built from different quartets are suppressed, because the two matrix elements share only a vanishing fraction of microscopic phase variables. This is the precise sense in which the present construction approaches the SYK limit.

The same phase randomization that Gaussianizes the active sector also suppresses the off-diagonal overlaps introduced above. The refined overlap matrix approaches the identity as $M^{-1/2}$, up to the trivial $O(M^{-1})$ correction to the diagonal in the equal-cell discretization. Orthogonalization therefore changes a typical nonzero coupling only perturbatively at large $M$, while seed-basis zeros are lifted only into parametrically weak tails. The large-$M$ theory is thus an asymptotically canonical \emph{sparse} SYK network where the active sector follows the Gaussian SYK law.

A useful reference point is an ideal IID-patch ensemble in which each parent orbital is replaced by $M$ statistically independent localized pieces whose centers, widths, and phases are all IID random variables. In that benchmark, Gaussian one-point statistics and asymptotically negligible connected inter-coupling correlations follow rather directly at large $M$ because each matrix element is built from many independent microscopic amplitudes. Closely related ideas underlie several earlier proposals to engineer SYK couplings from strongly irregular wave functions or from a large number of auxiliary microscopic channels~\cite{PikulinFranz2017,ChewEssinAlicea2017,ChenIlanDeJuanPikulinFranz2018,DanshitaHanadaTezuka2017}. The point of the present section is stronger: we will show that even after imposing substantial correlations in the centers and widths - namely, the correlations required when the microscopic pieces are forced to form a partition of one parent localization interval - the large-$M$ limit still produces Gaussian, asymptotically uncorrelated disorder for generic couplings.

It is also useful to state what kind of single-particle Hamiltonian can underlie the two partition ensembles. A strictly one-channel 1D Hamiltonian with a real scalar potential, $H=-\partial_x^2/2m+U(x)$, is not enough: with open boundaries its eigenfunctions can be chosen real, so it cannot generate the independently phased microscopic pieces used below. The natural microscopic setting is instead a multicomponent or higher-dimensional parent problem whose low-energy states are localized along an effectively one-dimensional manifold~\cite{WeiSedrakyan2023}. A minimal continuum form is
\begin{equation}
H_0=\int dx\,\hat\Psi^\dagger(x)\left[-\frac{1}{2m}\big(\partial_x-i\mathcal A(x)\big)^2+U(x)+W(x)\right]\hat\Psi(x),
\end{equation}
where $\hat\Psi$ is an $n_{\rm ch}$-component fermion field, $U(x)$ localizes the coarse envelope, and the matrix-valued terms $\mathcal A(x)$ and $W(x)$ mix internal channels and generate complex phase texture on a microscopic scale $\xi$. An equivalent lattice realization is a block Anderson chain with several orbitals per cell and complex random intra- and inter-cell hopping. If the weight inside one localization volume is roughly uniform from cell to cell, coarse graining gives the equal-cell ensemble discussed below. If it is redistributed into domains of irregular size, the same localized eigenstate is better described by the random partition ensemble that will be discussed and studied in the next subsections.

 We first introduce a \emph{random partition ensemble}, in which each parent localization interval is cut by random internal boundaries into contiguous subintervals. We then introduce an \emph{equal-cell partition ensemble}, obtained by freezing those internal widths to a regular mesh of size $\xi$. The second construction is thus a constrained, equal-width limit of the first, not a separate mechanism. 


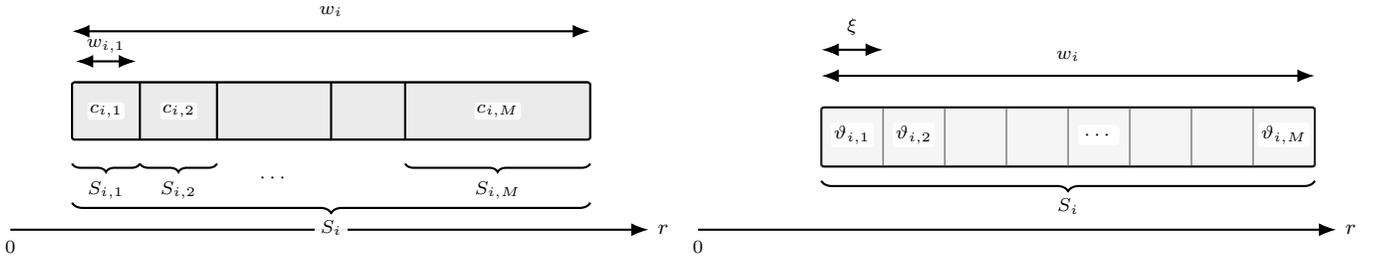
\begin{figure*}[t]
    \centering
    \tikzset{
        patchFill/.style={fill=black!8},
        patchLine/.style={draw=black,thick,rounded corners=1pt},
        envFill/.style={fill=black!4},
        envLine/.style={draw=black,thick,rounded corners=1pt},
        pixLine/.style={draw=black!70,thin},
        dim/.style={line width=0.8pt},
        lab/.style={fill=white,fill opacity=0.92,text opacity=1,text=black,inner sep=2.2pt,rounded corners=1.2pt},
        labS/.style={lab,font=\scriptsize},
    }

    \begin{subfigure}[t]{0.49\textwidth}
        \centering
        \begin{tikzpicture}[x=0.82cm,y=1.00cm,>=Latex, every node/.style={font=\footnotesize}]
            \draw[->,line width=0.8pt] (0,-0.22) -- (10.35,-0.22) node[right] {$r$};
            \node[labS,anchor=north,yshift=-2pt] at (0,-0.22) {$0$};

            \fill[envFill] (1.00,0.98) rectangle (9.40,1.74);
            \draw[envLine] (1.00,0.98) rectangle (9.40,1.74);

            \foreach \xA/\xB in {1.00/2.10,2.10/3.35,3.35/5.20,5.20/6.40,6.40/9.40} {
                \fill[patchFill] (\xA,0.98) rectangle (\xB,1.74);
            }
            \foreach \x in {2.10,3.35,5.20,6.40} {
                \draw[patchLine] (\x,0.98) -- (\x,1.74);
            }
            \draw[envLine] (1.00,0.98) rectangle (9.40,1.74);

            \draw[decorate,decoration={brace,mirror,amplitude=3.2pt},line width=0.8pt]
                (1.00,0.66) -- (2.10,0.66)
                node[midway,labS,below=4pt] {$S_{i,1}$};
            \draw[decorate,decoration={brace,mirror,amplitude=3.2pt},line width=0.8pt]
                (2.10,0.66) -- (3.35,0.66)
                node[midway,labS,below=4pt] {$S_{i,2}$};
            \draw[decorate,decoration={brace,mirror,amplitude=3.2pt},line width=0.8pt]
                (6.40,0.66) -- (9.40,0.66)
                node[midway,labS,below=4pt] {$S_{i,M}$};
            \node[labS] at (4.28,0.47) {$\cdots$};

            \draw[decorate,decoration={brace,mirror,amplitude=4pt},line width=0.8pt]
                (1.00,0.14) -- (9.40,0.14)
                node[midway,labS,below=4pt] {$S_i$};

            \draw[<->,dim] (1.06,2.02) -- (2.04,2.02);
            \node[labS,inner sep=0.8pt] at (1.55,2.23) {$w_{i,1}$};
            \draw[<->,dim] (1.00,2.42) -- (9.40,2.42)
                node[midway,labS,above=3pt] {$w_i$};

            \node[labS,inner sep=0.8pt] at (1.55,1.36) {$c_{i,1}$};
            \node[labS,inner sep=0.8pt] at (2.73,1.36) {$c_{i,2}$};
            \node[labS,inner sep=0.8pt] at (7.90,1.36) {$c_{i,M}$};
        \end{tikzpicture}
        \caption{\capj{random partition ensemble: the parent interval $S_i$ is cut by $M-1$ random internal boundaries into $M$ contiguous subintervals $S_{i,a}$. The small braces explicitly mark the first, second, and last subintervals. Their widths satisfy $\sum_a w_{i,a}=w_i$, so the widths and centers are correlated; only the phases are IID.}}
        \label{fig:I_schematic}
    \end{subfigure}
    \hfill
    \begin{subfigure}[t]{0.49\textwidth}
        \centering
        \begin{tikzpicture}[x=0.82cm,y=1.05cm,>=Latex, every node/.style={font=\footnotesize}]
            \draw[->,line width=0.8pt] (0,0) -- (10.35,0) node[right] {$r$};
            \node[labS,anchor=north,yshift=-2pt] at (0,0) {$0$};

            \fill[envFill] (2.0,0.80) rectangle (10.0,1.55);
            \draw[envLine] (2.0,0.80) rectangle (10.0,1.55);
            \draw[decorate,decoration={brace,mirror,amplitude=4pt},line width=0.8pt]
                (2.0,0.62) -- (10.0,0.62)
                node[midway,labS,below=4pt] {$S_i$};

            \foreach \k in {0,...,8} {
                \draw[pixLine] (2.0+1.0*\k,0.80) -- (2.0+1.0*\k,1.55);
            }

            \draw[<->,dim] (2.0,1.95) -- (10.0,1.95)
                node[midway,labS,above=3pt] {$w_i$};
            \draw[<->,dim] (2.0,2.28) -- (3.0,2.28)
                node[midway,labS,above=3pt] {$\xi$};

            \node[labS,inner sep=1.2pt] at (2.50,1.20) {$\vartheta_{i,1}$};
            \node[labS,inner sep=1.2pt] at (3.50,1.20) {$\vartheta_{i,2}$};
            \node[labS] at (6.50,1.20) {$\cdots$};
            \node[labS,inner sep=1.2pt] at (9.50,1.20) {$\vartheta_{i,M}$};
        \end{tikzpicture}
        \caption{\capj{Equal-cell partition ensemble: the same parent interval is resolved into equal cells of size $\xi$. This is the equal-width, reduced-randomness limit of the random partition ensemble, obtained by freezing the internal geometry and retaining only the phase disorder.}}
        \label{fig:II_schematic}
    \end{subfigure}

    \caption{\capj{Two microscopic routes to Gaussianization. Panel (a) shows the random partition ensemble. Panel (b) shows the equal-cell partition ensemble, which is the equal-width limit of panel (a). In both cases a single nonzero coupling is built from many microscopic contributions, and the central-limit mechanism of Eq.~(\ref{eq:CLT_sum}) drives the active sector toward the SYK Gaussian law.}}
    \label{fig:true_syk_schematics}
\end{figure*}

\subsection{random partition ensemble}
\label{sec:true_syk_random_partition}

We start from the same parent support $S_i=[c_i-w_i/2,\,c_i+w_i/2]$ introduced in Sec.~\ref{sec:geometric_construction}. Inside each parent interval we draw $M-1$ internal boundaries, order them, and define
\begin{eqnarray}
\label{eq:partition_boundaries}
&c_i-\frac{w_i}{2}\equiv b_{i,0}<b_{i,1}<\cdots<b_{i,M}\equiv c_i+\frac{w_i}{2},\nonumber\\
&S_{i,a}\equiv[b_{i,a-1},b_{i,a}].
\end{eqnarray}
Thus the $S_{i,a}$ form a genuine partition of $S_i$: they are contiguous, never overlap, leave no empty gaps, and satisfy $\sum_{a=1}^{M}w_{i,a}=w_i$, where $w_{i,a}=|S_{i,a}|$. If the internal boundaries are obtained as the ordered values of $M-1$ IID points uniformly distributed in $S_i$, then the normalized widths $w_{i,a}/w_i$ are Dirichlet distributed, with $\langle w_{i,a}/w_i\rangle=1/M$ and covariance $\mathrm{Cov}(w_{i,a}/w_i,w_{i,b}/w_i)=-1/[M^2(M+1)]$ for $a\neq b$. The centers are correlated as well,
\begin{equation}
\label{eq:partition_centers}
c_{i,a}=c_i-\frac{w_i}{2}+\sum_{m<a}w_{i,m}+\frac{w_{i,a}}{2},
\end{equation}
so once one subinterval is widened, all subsequent centers are shifted. The random partition ensemble is therefore random but emphatically \emph{not} an IID patch model.

This distinction is physically important. A wide subinterval necessarily depletes the remaining available length; two subintervals can never overlap, and the absence of gaps ties the entire internal geometry to one cumulative set of boundaries. In other words, the microscopic pieces do not float independently inside $S_i$ but tile it exactly. The phases, however, remain independent. The ensemble is thus hybrid: the geometry is correlated by construction, whereas the phases are IID. In this section, we will precisely test whether these geometric correlations obstruct the emergence of SYK disorder. The numerical answer will be no.

The orbital is defined as
\begin{eqnarray}
\label{eq:multipatch_wf}
&&\Psi_i(r)=\frac{1}{\sqrt{M}}\sum_{a=1}^{M}\psi_{i,a}(r),\nonumber\\
&&\psi_{i,a}(r)=\frac{e^{i\varphi_{i,a}}}{\sqrt{w_{i,a}}}\,\theta\inb({w_{i,a}-2|r-c_{i,a}|}),
\end{eqnarray}
with $\varphi_{i,a}$ IID uniform on $[0,2\pi)$. Because the supports $S_{i,a}$ are disjoint, the prefactor $1/\sqrt{M}$ keeps the normalization exact, $\int_0^Ldr\,|\Psi_i(r)|^2=1$, up to the same negligible sample-edge corrections discussed in Sec.~\ref{sec:geometric_construction}. Eq.~(\ref{eq:multipatch_wf}) is therefore the natural partition-based refinement of the rectangular orbital of Sec.~\ref{sec:geometric_construction}: each parent state is now endowed with internal structure while preserving its overall support and normalization.

The unsymmetrized matrix element is defined by the same projection formula as Eq.~(\ref{eq:Jbar_def}), but with $\psi_i\mapsto\Psi_i$,
\begin{equation}
\label{eq:multipatch_Jbar}
\bar J_{ij}^{kl}=
\int dr_1dr_2\,\Psi_i^\ast(r_1)\Psi_j^\ast(r_2)\Psi_k(r_1)\Psi_l(r_2)\,V(r_2-r_1).
\end{equation}
Expanding Eq.~(\ref{eq:multipatch_wf}) gives
\begin{equation}
\label{eq:multipatch_Jbar_sum}
\bar J_{ij}^{kl}=\frac{1}{M^2}\sum_{a,b,c,d=1}^{M}\bar J_{(i,a)(j,b)}^{(k,c)(l,d)},
\end{equation}
with patch-level contributions
\begin{equation}
\label{eq:patch_coupling_def}
\bar J_{(i,a)(j,b)}^{(k,c)(l,d)}=
\int dr_1dr_2\,
\psi_{i,a}^\ast(r_1)\psi_{j,b}^\ast(r_2)\psi_{k,c}(r_1)\psi_{l,d}(r_2)\,V(r_2-r_1).
\end{equation}
Eq.~(\ref{eq:multipatch_Jbar_sum}) has exactly the structure needed for Gaussianization, but with correlated amplitudes rather than independent ones. Changing one internal boundary simultaneously changes two neighboring widths, shifts the centers of all subsequent cells through Eq.~(\ref{eq:partition_centers}), and, through $\sum_a w_{i,a}=w_i$, redistributes length across the entire parent interval. Nearby microscopic amplitudes are therefore correlated already at the geometric level.

These correlations do \emph{not} invalidate the basic mechanism. For distinct orbital labels $i,j,k,l$, each patch-level term carries the phase factor $e^{i(\varphi_{k,c}+\varphi_{l,d}-\varphi_{i,a}-\varphi_{j,b})}$, so phase averaging still gives $\langle \bar J_{ij}^{kl}\rangle=0$ and removes all off-diagonal terms in the variance,
\begin{equation}
\label{eq:multipatch_variance}
\left\langle \left|\bar J_{ij}^{kl}\right|^2\right\rangle
=
\frac{1}{M^4}\sum_{a,b,c,d=1}^{M}
\left\langle \left|\bar J_{(i,a)(j,b)}^{(k,c)(l,d)}\right|^2\right\rangle.
\end{equation}
The role of the partition-induced correlations is therefore not to destroy self-averaging but to reshape the finite-$M$ crossover. They enhance the weight of rare quartets for which only a small number of microscopic overlap cells contributes appreciably, thereby feeding the near-zero sector seen numerically below. Once many partition cells participate inside a typical overlap region, however, the continuous part of the coupling distribution is still driven toward the SYK Gaussian form.

This is the main physical message of the random partition ensemble. Complete microscopic independence is unnecessary. What matters is that a typical nonzero matrix element receives contributions from many pieces with independent phases and finite variance, even if the amplitudes of those pieces remain geometrically correlated. The equal-cell construction discussed next makes that conclusion especially transparent by freezing the width fluctuations while leaving the phase disorder intact.

\subsection{Equal-cell partition ensemble}
\label{sec:true_syk_equal_cell}

We now turn to the equal-cell partition ensemble, which is the equal-width limit of the random partition ensemble. We partition the sample into microscopic cells
\begin{eqnarray}
\label{eq:pixel_partition}
&&P_a\equiv[(a-1)\xi,a\xi],
\qquad
r_a\equiv\left(a-\frac12\right)\xi,\nonumber\\
 &&a=1,\dots,M_{\rm tot}\equiv L/\xi,
\end{eqnarray}
and denote by
$\Lambda_i\equiv\{a\,|\,P_a\subset S_i\}\subset\{1,\dots,M_{\rm tot}\}$
the set of cell labels fully contained in the parent interval $S_i$. In other words, $\Lambda_i$ is the list of microscopic cells occupied by orbital $i$. This is precisely the partition construction above with all internal widths frozen to $\xi$, up to the harmless $O(\xi)$ mismatch at the two ends of $S_i$. The equal-cell ensemble should therefore be read as a controlled deformation of the random partition ensemble.

The orbital becomes
\begin{equation}
\label{eq:randomphase_wf}
\Psi_i(r)=\frac{1}{\sqrt{w_i}}\sum_{a\in\Lambda_i} e^{i\vartheta_{i,a}}\,\theta\inb({\xi-2|r-r_a|}),
\end{equation}
with IID phases $\vartheta_{i,a}\in[0,2\pi)$. Thus the label $a\in\Lambda_i$ simply selects one occupied microscopic cell inside orbital $i$. It is useful to view Eq.~(\ref{eq:randomphase_wf}) as Eq.~(\ref{eq:multipatch_wf}) with all occupied subintervals set equal to $\xi$. Since $|\Psi_i(r)|=1/\sqrt{w_i}$ on each occupied cell, the normalization remains $\int_0^Ldr\,|\Psi_i(r)|^2=1+O(\xi/w_i)$. The correction comes only from the boundary cells discarded when defining $\Lambda_i$. In the constant-width ensemble of Sec.~\ref{sec:geometric_construction}, all orbitals contain the same number $M\simeq w_i/\xi$ of microscopic cells, and this is the quantity denoted by $M$ in the numerical plots below. For variable parent widths one simply has $M_i\equiv|\Lambda_i|\simeq w_i/\xi$.

At this point, the orthogonalization becomes explicit. The overlap matrix of the refined orbitals is
\begin{eqnarray}
&&\!\!\!\!\!\Omega^{(M)}_{ij}\equiv\int_0^Ldr\,\Psi_i^*(r)\Psi_j(r)
=\frac{\xi}{\sqrt{w_iw_j}}\sum_{a\in\Lambda_i\cap\Lambda_j}e^{i(\vartheta_{j,a}-\vartheta_{i,a})},\nonumber\\
&&\!\!\!\!\!\Big\langle |\Omega^{(M)}_{ij}|^2\Big\rangle_{\vartheta}
=\frac{\xi^2|\Lambda_i\cap\Lambda_j|}{w_iw_j}\simeq \frac{\xi\,\ell_{ij}}{w_iw_j},
\qquad i\neq j.
\label{eq:overlap_matrix_largeM}
\end{eqnarray}
Here $\Lambda_i\cap\Lambda_j$ is the set of microscopic cells shared by the two parent intervals. Since $\ell_{ij}\le \min(w_i,w_j)$, one has $\langle |\Omega^{(M)}_{ij}|^2\rangle \le \xi/\max(w_i,w_j)$. For constant widths this gives $|\Omega^{(M)}_{ij}|_{\rm typ}=O(M^{-1/2})$ with $M\simeq w/\xi$, and the same scaling holds more generally in terms of the number of microscopic cells inside the overlap region. The random partition ensemble behaves the same way, because its overlap matrix is again a sum of many independently phased microscopic pieces. Thus $\Omega^{(M)}=I+O(M^{-1/2})$ after trivial diagonal normalization. The transformation of Eq.~(\ref{eq:lowdin_orbitals}) therefore changes a typical nonzero quartic coupling only by a relative $O(M^{-1/2})$, while strict seed zeros are lifted only into parametrically weak tails. At any fixed threshold that does not scale to zero with $M$, those tails remain subleading in the graph analysis. The large-$M$ limit therefore yields canonical fermions together with Gaussian statistics on the active sector.

Because the wave function is constant on each occupied cell, the interaction integral reduces to a double sum over cell pairs. Writing $\Lambda_{ik}\equiv \Lambda_i\cap\Lambda_k$ and $\Lambda_{jl}\equiv \Lambda_j\cap\Lambda_l$, one obtains
\begin{eqnarray}
\label{eq:randomphase_Jbar_sum}
&&\bar J_{ij}^{kl}=
\frac{1}{\sqrt{w_iw_jw_kw_l}}\\
&&\times\sum_{a\in\Lambda_{ik}}\sum_{b\in\Lambda_{jl}}
\xi^2\,V(r_b-r_a)
\,e^{i(\vartheta_{k,a}+\vartheta_{l,b}-\vartheta_{i,a}-\vartheta_{j,b})}.\nonumber
\end{eqnarray}
This is again of the form \refe{CLT_sum}, now with $M_{\rm eff}\sim |\Lambda_{ik}|\,|\Lambda_{jl}|\sim (\ell_{ik}\ell_{jl})/\xi^2$, where $\ell_{ab}\equiv |S_{ab}|$ uses the overlap interval $S_{ab}$ already defined in Sec.~\ref{sec:geometric_construction}. Thus $M_{\rm eff}$ diverges at fixed parent overlaps as $\xi\to0$. The equal-cell partition ensemble is therefore the cleanest demonstration that full geometric randomness is not the essential ingredient: even after the widths are frozen, the same Gaussian limit emerges because each matrix element still samples many independently phased microscopic terms.

The variance follows from the same diagonal-phase argument as above. Expanding $|\bar J_{ij}^{kl}|^2$ and averaging over the IID phases removes all off-diagonal terms with $(a,b)\neq(a',b')$. Approximating the remaining diagonal sum by the corresponding double integral yields
\begin{eqnarray}
\label{eq:randomphase_variance_general}
\left\langle \left|\bar J_{ij}^{kl}\right|^2\right\rangle_{\vartheta}
=
\frac{\xi^2}{w_iw_jw_kw_l}
\int_{S_{ik}}dr_1\int_{S_{jl}}dr_2\,V(r_2-r_1)^2\nonumber\\
+O\!\left(\frac{1}{M_{\rm eff}}\right).
\end{eqnarray}
The scaling is physically transparent: each surviving diagonal cell pair contributes an area factor $\xi^2$, while the number of such pairs is already encoded in the two overlap lengths entering the integral. After phase averaging, the variance therefore scales as $\xi^2$ and the rms coupling scales linearly with $\xi$.


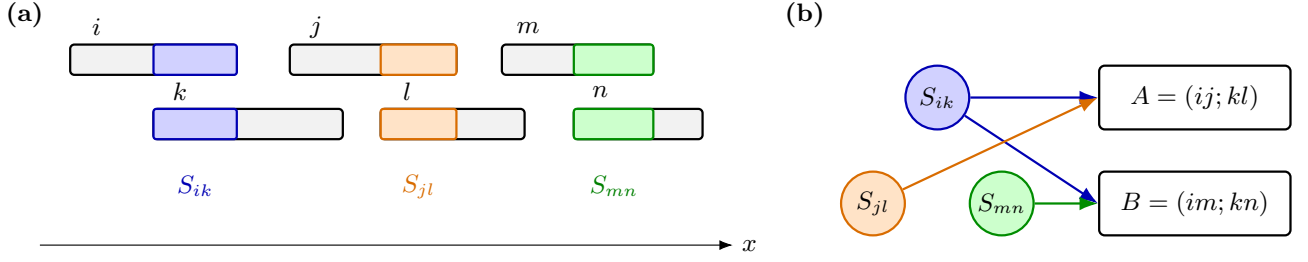
\begin{figure*}[t]
\centering
\resizebox{0.96\textwidth}{!}{%
\begin{tikzpicture}[x=0.95cm,y=0.95cm,>=Latex,
    orbit/.style={draw, thick, rounded corners=1.5pt, fill=gray!10},
    sharedov/.style={draw=blue!70!black, fill=blue!18, thick, rounded corners=1pt},
    ovA/.style={draw=orange!85!black, fill=orange!22, thick, rounded corners=1pt},
    ovB/.style={draw=green!55!black, fill=green!20, thick, rounded corners=1pt},
    setnode/.style={circle, draw, thick, minimum size=8mm, inner sep=1pt},
    coup/.style={draw, thick, rounded corners=2pt, minimum width=24mm,
                 minimum height=8mm, align=center, fill=white}
]

\node[font=\bfseries] at (-0.20,3.35) {(a)};
\draw[->] (0,0.30) -- (9.15,0.30) node[right] {$x$};

\draw[orbit] (0.40,2.55) rectangle (2.60,2.95);
\node at (0.75,3.18) {$i$};
\draw[orbit] (1.50,1.70) rectangle (4.00,2.10);
\node at (1.85,2.33) {$k$};
\draw[sharedov] (1.50,2.55) rectangle (2.60,2.95);
\draw[sharedov] (1.50,1.70) rectangle (2.60,2.10);
\node[blue!70!black] at (2.05,1.08) {$S_{ik}$};

\draw[orbit] (3.30,2.55) rectangle (5.50,2.95);
\node at (3.65,3.18) {$j$};
\draw[orbit] (4.50,1.70) rectangle (6.40,2.10);
\node at (4.85,2.33) {$l$};
\draw[ovA] (4.50,2.55) rectangle (5.50,2.95);
\draw[ovA] (4.50,1.70) rectangle (5.50,2.10);
\node[orange!85!black] at (5.00,1.08) {$S_{jl}$};

\draw[orbit] (6.10,2.55) rectangle (8.10,2.95);
\node at (6.45,3.18) {$m$};
\draw[orbit] (7.05,1.70) rectangle (8.75,2.10);
\node at (7.40,2.33) {$n$};
\draw[ovB] (7.05,2.55) rectangle (8.10,2.95);
\draw[ovB] (7.05,1.70) rectangle (8.10,2.10);
\node[green!50!black] at (7.58,1.08) {$S_{mn}$};


\node[font=\bfseries] at (10.10,3.35) {(b)};

\node[setnode, fill=blue!18, draw=blue!70!black]      (Sik) at (11.85,2.25) {$S_{ik}$};
\node[setnode, fill=orange!22, draw=orange!85!black]  (Sjl) at (11.00,0.85) {$S_{jl}$};
\node[setnode, fill=green!20, draw=green!55!black]    (Smn) at (12.70,0.85) {$S_{mn}$};

\node[coup] (A) at (15.25,2.25) {$A=(ij;kl)$};
\node[coup] (B) at (15.25,0.85) {$B=(im;kn)$};

\draw[->, thick, blue!70!black]   (Sik) -- (A.west);
\draw[->, thick, blue!70!black]   (Sik) -- (B.west);
\draw[->, thick, orange!85!black] (Sjl) -- (A.west);
\draw[->, thick, green!55!black]  (Smn) -- (B.west);


\end{tikzpicture}%
}
\caption{Representative example of two couplings sharing one microscopic leg.
(a) Three real-space overlap sets built from localized orbitals: the common overlap $S_{ik}$ (blue), the overlap $S_{jl}$ entering $A=(ij;kl)$ (orange), and the overlap $S_{mn}$ entering $B=(im;kn)$ (green).
(b) Schematic structure of the two couplings. Both $A$ and $B$ contain the same microscopic overlap set $S_{ik}$, represented by the shared blue leg, while their second overlap sets are different: $S_{jl}$ for $A$ and $S_{mn}$ for $B$. This is the simplest nontrivial example of two distinct couplings that are not fully independent because they share one microscopic leg.}
\label{fig:shared_microscopic_leg}
\end{figure*}


The one-point Gaussianization derived above is only part of the SYK limit. One must also control connected correlations between distinct active couplings. In the equal-cell ensemble, this can be estimated explicitly at fixed parent geometry. For a quartet $A\equiv(ij;kl)$, Eq.~(\ref{eq:randomphase_Jbar_sum}) can be written as

\begin{equation}
\bar J_A=\sum_{a\in\Lambda_{ik}}\sum_{b\in\Lambda_{jl}}u^{(A)}_{ab}X_a^{(ik)}Y_b^{(jl)},
\quad
u^{(A)}_{ab}\equiv \frac{\xi^2V(r_b-r_a)}{\sqrt{w_iw_jw_kw_l}},
\label{eq:J_corr_factorized}
\end{equation}
where $X_a^{(ik)}\equiv e^{i(\vartheta_{k,a}-\vartheta_{i,a})}$ and $Y_b^{(jl)}\equiv e^{i(\vartheta_{l,b}-\vartheta_{j,b})}$ are unit-modulus random phases with zero mean. Consider another coupling $B\equiv(mn;pq)$. If the four orbital pairs $(i,k)$, $(j,l)$, $(m,p)$, and $(n,q)$ are all distinct, then the two couplings depend on disjoint phase families. Therefore
\begin{equation}
\langle \bar J_A \bar J_B^*\rangle_{\vartheta}=\langle \bar J_A \bar J_B\rangle_{\vartheta}=0,
\qquad
\langle |\bar J_A|^2|\bar J_B|^2\rangle_{c,\vartheta}=0,
\label{eq:disjoint_corr_zero}
\end{equation}
exactly, where $\langle XY\rangle_c\equiv \langle XY\rangle-\langle X\rangle\langle Y\rangle$. Thus generic couplings built from disjoint quartets are already independent at the level of phase averaging.

The first nontrivial case is when two couplings share one microscopic leg. A representative example, shown in Fig.~\ref{fig:shared_microscopic_leg}, is $A=(ij;kl)$ and $B=(im;kn)$, which share the same $ik$ overlap set but have different second overlap sets. In that case $\langle \bar J_A \bar J_B^*\rangle_{\vartheta}=0$ still, while the connected intensity correlation becomes
\begin{equation}
\langle |\bar J_A|^2|\bar J_B|^2\rangle_{c,\vartheta}
=
\sum_{a\neq a'\in\Lambda_{ik}} C^{A}_{aa'}C^{B}_{a'a},
\label{eq:shared_leg_corr}
\end{equation}
\begin{equation}
C^{A}_{aa'}\equiv \sum_{b\in\Lambda_{jl}}u^{(A)}_{ab}u^{(A)\,*}_{a'b},
\quad
C^{B}_{aa'}\equiv \sum_{d\in\Lambda_{mn}}u^{(B)}_{ad}u^{(B)\,*}_{a'd}.
\label{eq:shared_leg_corr_blocks}
\end{equation}
For typical widths $w_\alpha\sim w\sim M\xi$ and bounded $V$ on the support scale, one has $u^{(A)}_{ab},u^{(B)}_{ad}=O(M^{-2})$, hence $C^{A}_{aa'},C^{B}_{aa'}=O(M^{-3})$. Since the sum in Eq.~(\ref{eq:shared_leg_corr}) contains $O(M^2)$ terms,
\begin{equation}
\langle |\bar J_A|^2|\bar J_B|^2\rangle_{c,\vartheta}=O(M^{-4}).
\label{eq:shared_leg_scaling}
\end{equation}
This is the same order as $\langle |\bar J_A|^2\rangle_{\vartheta}\langle |\bar J_B|^2\rangle_{\vartheta}$, so pairs of couplings that share an entire microscopic leg remain correlated. They are, however, nongeneric: among $O(N^8)$ ordered pairs of quartic couplings, only $O(N^6)$ share a full orbital pair, so their fraction is $O(N^{-2})$. The random partition ensemble obeys the same $M$-counting, because the phases are again independent while the partition constraint changes only the amplitudes. After antisymmetrization and orthogonalization, these estimates remain unchanged up to the already established $O(M^{-1/2})$ corrections. In this sense the orthogonalized large-$M$ theory reproduces the SYK one-point law on the active sector and SYK-like inter-coupling statistics for the overwhelming majority of distinct couplings.


For the short-range interaction $V_u(r)=\theta(r_0-|r|)$ one has $V_u^2=V_u$, so the interval--interval integral in \refe{randomphase_variance_general} is controlled by the same functional $\mathcal{I}_{V_u}$ introduced in \refe{JV_def}. For the regularized Coulomb interaction $V_c(r)=1/\max(|r|,r_0)$, the squared kernel is $V_c(r)^2=1/\max(|r|,r_0)^2$, and the corresponding second primitive may be chosen as
\begin{equation}
\label{eq:Ic2_def}
I_{c,2}(r)=
\theta(r_0-|r|)\frac{r^2}{2r_0^2}
+\theta(|r|-r_0)\left[-\ln\!\left(\frac{|r|}{r_0}\right)+2\frac{|r|}{r_0}-\frac{3}{2}\right].
\end{equation}
Here, the inner-core branch and the outer-tail branch of $I_{c,2}(r)$ agree in both value and first derivative at $r=\pm r_0$. In particular, $I_{c,2}(r)$ has no cusp at the matching point. The only nonanalyticity is a jump in the second derivative, exactly where the kernel itself crosses over from the flat core $1/r_0^2$ to the tail $1/r^2$. Equivalently, $I_{c,2}''(r)=V_c(r)^2$ piecewise, so the endpoint formula \refe{JV_endpoints} can be applied directly without any extra boundary contribution. Eq.~(\ref{eq:randomphase_variance_general}) also shows that the rms coupling scales linearly with $\xi$. One may therefore impose the SYK normalization $\langle |J|^2\rangle=\mathcal{I}^2/N^3$ either by tuning the microscopic cell size or, equivalently, by a global rescaling $V\to gV$.

The equal-cell partition ensemble is useful precisely because it removes the width fluctuations of the general partition construction while leaving the phase-randomization mechanism intact. Any remaining difference between the two ensembles can therefore be attributed directly to partition-induced geometric correlations. This makes the numerical comparison especially sharp.

\subsection{Numerical evidence for Gaussianization}
\label{sec:true_syk_numerics}

\begin{figure*}
    \centering
    \begin{subfigure}[t]{.4\linewidth}
        \centering
        \includegraphics[width=\textwidth]{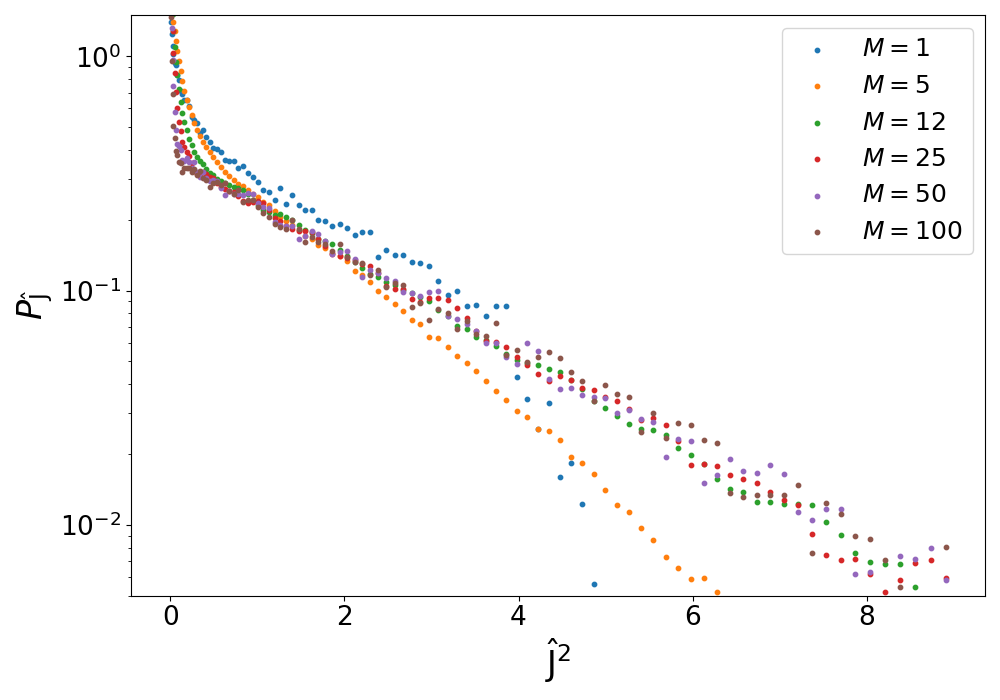}
        \includegraphics[width=\textwidth]{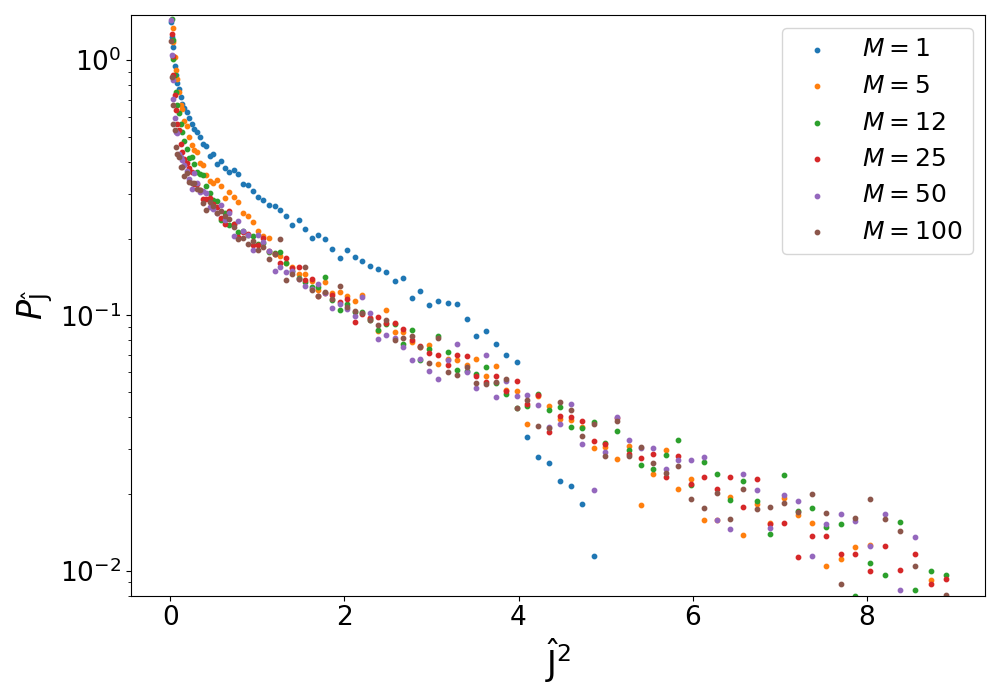}
        \caption{Active-sector distribution $P_{\hat J}$ of the normalized coupling $\hat J$, plotted as $\log P_{\hat J}$ versus $\hat J^2$ for several values of the internal resolution $M$. A straight segment in these coordinates is the signature of the Gaussian regular sector, since $P_{\hat J}\propto e^{-\kappa|\hat J|^2}$ appears linear as a function of $\hat J^2$.
            }
        \label{fig:hist_full}
    \end{subfigure}
    \squad 
    \begin{subfigure}[t]{.28\linewidth}
        \centering
        \includegraphics[width=\textwidth]{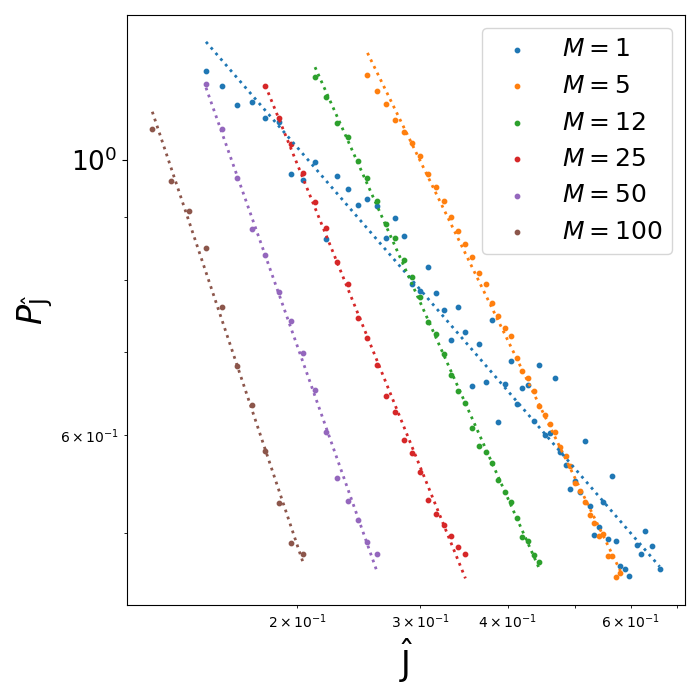}
        \includegraphics[width=\textwidth]{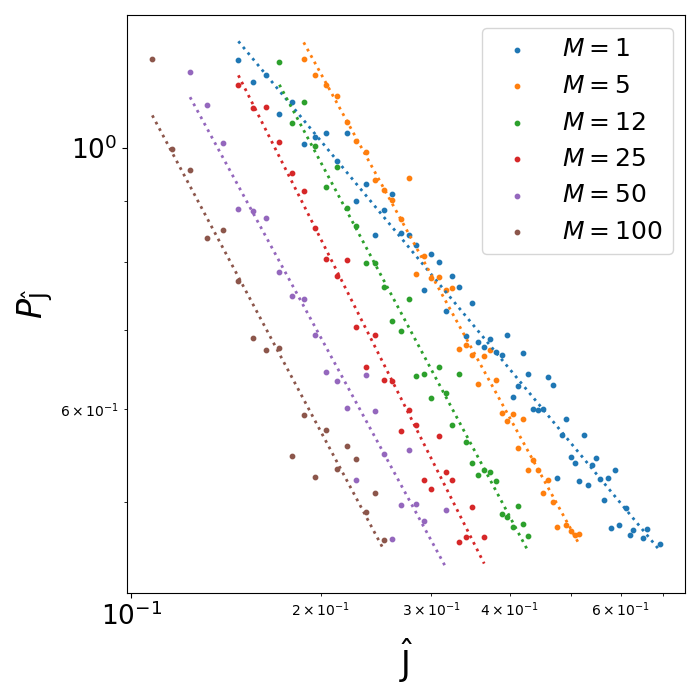}
        \caption{ Zoom of the singular near-zero sector, shown as $\log P_{\hat J}$ versus $\log \hat J$ for the same values of $M$. The near-linear behavior on these log-log axes shows the power-law decaying form of the singular sector close to $\hat J=0$.
          }
        \label{fig:hist_sing}
    \end{subfigure}
    \squad
    \begin{subfigure}[t]{.24\linewidth}
        \centering
        \includegraphics[width=\textwidth]{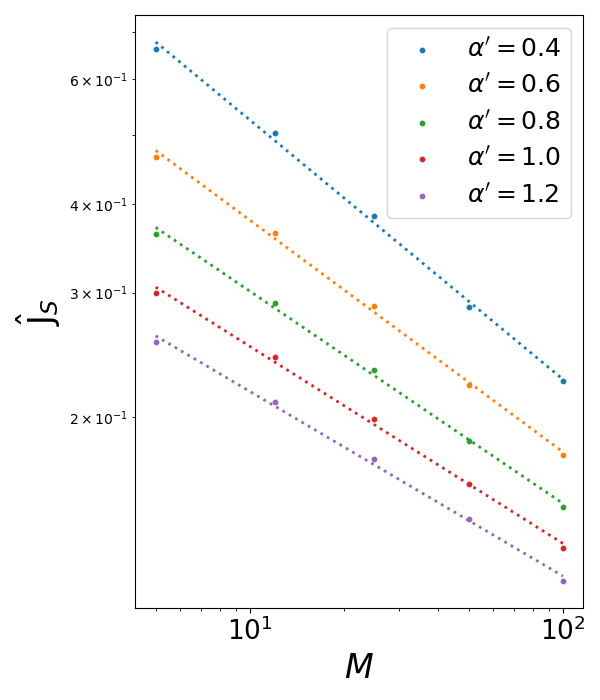}
        \includegraphics[width=\textwidth]{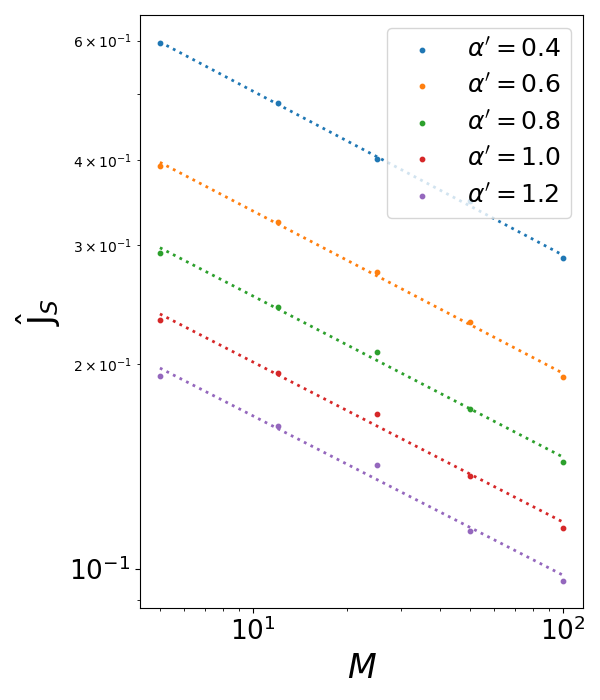}
        \caption{Width $\hat J_S$ of the singular near-zero sector, plotted against $M$ on log--log axes for several cutoff values $\alpha'$. The decrease of $\hat J_S$ with $M$ shows that the singular sector shrinks as the internal resolution increases.
          }
        \label{fig:hist_width}
    \end{subfigure}
    \caption{
    Numerical evidence for Gaussianization of the active sector. All histograms are constructed from the active sector, i.e., after removing the exact zeros and renormalizing the remaining distribution. The top row shows the random partition ensemble and the bottom row the equal-cell partition ensemble. As the internal resolution $M$ increases, the Gaussian regular sector grows, while the singular near-zero sector becomes narrower. Column (a) shows the semilog plot used to identify the Gaussian regular sector, column (b) zooms into the singular near-zero sector, and column (c) shows the decay of its width $\hat J_S$.
        }
    \label{fig:hist_all}
\end{figure*}


As in Sec.~\ref{sec:distJ}, the exact geometric zeros are separate from the continuous sector. We therefore construct the histograms in \reff{hist_all} from the \emph{active sector}, that is, from the full distribution after removing the point mass at $J=0$ and renormalizing the remainder.

The main numerical message of \reff{hist_full} and \reff{hist_sing} is quite robust. Increasing the internal resolution of a localized orbital does two things simultaneously: it expands a Gaussian regular sector at generic couplings, and it compresses a singular near-zero sector associated with rare quartets that still sample too few microscopic contributions. To make this visible, we normalize the physical couplings by their rms value,
\begin{equation}
\hat J\equiv \frac{J}{\sqrt{\langle |J|^2\rangle}},
\qquad
\int d^2\hat J\,P_{\hat J}(\hat J)=1,
\end{equation}
and decompose the resulting complex density as
\begin{equation}
\label{eq:hatJ_distr}
P_{\hat J}(\hat J)
=
A\,\frac{\kappa}{\pi}e^{-\kappa |\hat J|^2}
+(1-A)\,S^{(M)}(\hat J),
\end{equation}
where $S^{(M)}$ is a normalized singular contribution concentrated near $\hat J=0$.

The plots in \reff{hist_full} show why this form is natural. Once $M$ is moderately large, the histograms develop an extended interval that is strikingly linear when plotted as $\log P_{\hat J}$ versus $\hat J^2$. This is precisely the signature of the circular-Gaussian term in Eq.~(\ref{eq:hatJ_distr}). The fitted slopes give $\kappa\approx0.4$ for the random partition ensemble and $\kappa\approx0.3$ for the equal-cell partition ensemble. These values should be interpreted as the width of the Gaussian \emph{regular sector}, not of the full normalized distribution. If the singular part is sufficiently concentrated near the origin, its contribution to $\langle |\hat J|^2\rangle$ is negligible, and the normalization condition implies $A/\kappa\approx1$, hence $A\approx\kappa$. In other words, the Gaussian sector already carries essentially the entire second moment even though a finite fraction of the probability weight still remains trapped near $\hat J=0$.

An important physical point follows from \reff{hist_full}. Both ensembles develop the same regular Gaussian sector even though only one of them contains random internal widths. Exact independence of the microscopic geometry is therefore not required. The random partition ensemble retains the nontrivial correlations implied by the partition constraint, while the equal-cell partition ensemble freezes those correlations by hand. The fact that the same semilog-linear window emerges in both cases shows directly that the SYK limit is controlled by the multiplicity of independent \emph{random phases}, not by complete independence of the subinterval geometry.

The singular part in \reff{hist_all} has a simple origin. It comes from rare orbital quartets for which the effective number of microscopic contributions is still small at finite $M$. In the random partition ensemble this can happen in two ways. The parent overlaps $S_{ik}$ and $S_{jl}$ may already be small, or the random partition may place only a few subintervals inside the overlap region because the widths are constrained, nonoverlapping, and correlated. In the equal-cell partition ensemble the second effect is largely removed. There the near-zero weight is controlled mainly by small parent overlaps and by finite cell size. Comparing the two rows therefore isolates the role of partition-induced geometric correlations in the finite-$M$ crossover.
So this near-zero peak is not the exact $\delta$-peak of inactive quartets discussed before.  It is the finite-$M$ singular part of the active sector.

To quantify the near-zero sector, we define its characteristic width $\hat J_S(M)$ by the condition
\begin{equation}
(1-A)S^{(M)}(\hat J_S)=\alpha\,P_{\hat J}(\hat J_S),
\qquad 0<\alpha<1,
\end{equation}
which, under the same narrow-singular-sector assumption, is equivalent to
\begin{equation}
\label{eq:hist_sing}
P_{\hat J}(\hat J_S)=\frac{\kappa^2}{\pi(1-\alpha)} \equiv \alpha' \squad.
\end{equation}
In practice, we determine $\hat J_S(M)$  by finding where the near-zero part of the histogram $P_{\hat J}(\hat J)$ crosses a fixed reference level $\alpha'$.
\reff{hist_sing} is particularly useful for this.
One can see that the singular sectors of the distribution $P_{\hat J}$ are
well approximated by power law functions (the dotted lines on the figures),
and becomes narrower with an increasing value of $M$.
The functional forms of $\hat{J}_S(M)$ can be extracted from these plots
for an arbitrary parameter $\alpha'$,
and are shown to be of form $J_S(M)\propto M^{-\beta_s}$ (\reff{hist_width}),
 with  $\beta_s\approx0.3$ for the random partition ensemble and  $\beta_s\approx1/4$ for the equal-cell partition ensemble. These exponents are not universal. They simply describe how quickly the rare near-zero sector is pushed toward the origin as $M$ grows.

The physical picture is now straightforward. The Gaussian regular sector comes from matrix elements with large $M_{\rm eff}$, where many microscopic contributions with independent random phases are added together. In this regime, the central-limit mechanism drives the distribution toward a Gaussian form. By contrast, the singular sector comes from rare orbital quartets for which $M_{\rm eff}$ is still too small, so self-averaging has not yet developed. Comparing the two ensembles makes the role of partition-induced correlations very clear: they modify the finite-$M$ crossover and enhance the near-zero tail, but they do not prevent the emergence of the Gaussian SYK regular sector. In the same large-$M$ limit, two generic couplings built from different orbital quartets share only a negligible fraction of the underlying microscopic phase variables, so their connected cross-correlations are also suppressed. Couplings that share some orbital labels can still have subleading correlations at finite $M$, but these correlations no longer control the large-$M$ disorder statistics of generic couplings.

We therefore conclude that localized states in one dimension can produce the Gaussian SYK distribution for the nonzero couplings, provided that each localization volume contains enough internal phase structure. The microscopic pieces do not need to be fully independent: a correlated random partition is already sufficient. Replacing that random partition by equal cells does not change the large-$M$ limit, showing that the essential ingredient is phase self-averaging rather than fully random geometry. At the same time, the pattern of vanishing couplings is still set by the real-space overlap of the parent supports. Thus, the correct large-$M$ description is not a single fully connected SYK dot, but a sparse canonical SYK network whose densely connected components form the emergent SYK clusters discussed in the next section. 

From an experimental point of view, the internal resolution parameter should be interpreted as the number of incoherent phase domains, microscopic sites, or auxiliary channels contained within one localization volume. In the equal-cell picture, the basic requirement is
$M_{\rm eff}\sim \ell_{ik}\ell_{jl}/\xi^2 \gg 1$
for a typical quartet that gives a nonzero coupling. In addition, the platform must support overlapping but orthonormal fermionic orbitals whose internal wave functions carry complex phase structure.

\section{Graph representation and nucleation of SYK clusters}
\label{sec:graph_mapping}


In the previous sections, we established two complementary parts of the same one-dimensional construction. In Sec.~\ref{sec:correlated_disorder}, we showed how the overlap geometry of localized orbitals determines the basic disorder structure of the couplings. In Sec.~\ref{sec:true_syk}, we then showed that additional internal phase structure can drive the nonzero couplings toward Gaussian SYK statistics. A separate and equally important question is which of these couplings are strong enough to hybridize the orbitals into a common interacting cluster. Because every matrix element is ultimately generated by real-space overlap, the interaction tensor remains sparse at any finite population $N$, even in the Gaussianized regime.

In this section we focus on the graph formed by the strong couplings, that is, by keeping only matrix elements whose magnitude exceeds a fixed threshold. Because our goal here is to understand how dense interacting clusters form in pair space, rather than to revisit the detailed disorder law of the couplings, it is sufficient, and numerically much less expensive, to work with the $M=1$ tensor of Sec.~\ref{sec:correlated_disorder}. At $M=1$, the couplings are still the correlated and generally non-Gaussian couplings discussed in Sec.~\ref{sec:correlated_disorder}; they are not yet the canonical SYK couplings. The question addressed in the present section is different: we ask which scattering channels are connected strongly enough to build a cluster with nearly all-to-all mixing.

Sections~\ref{sec:correlated_disorder} and~\ref{sec:true_syk} already show how random internal phases drive the nonzero couplings toward the SYK distribution at large $M$. At the same time, Eq.~(\ref{eq:overlap_matrix_largeM}) shows that, in this regime, orthogonalization changes a typical strong edge only by a relative correction of order $M^{-1/2}$ and turns exact seed zeros into parametrically weak tails. Therefore, if the edge threshold is kept finite as $M\to\infty$, these tails remain too small to create a significant number of new strong links. In this sense, the strong-link connectivity, and hence the nucleation, growth, and merger of clusters, is the same to leading order in the $M=1$ and large-$M$ theories. As $M$ becomes large, what mainly changes is the statistics of the nonzero couplings on a given geometric network, not the network itself. For this reason, the $M=1$ calculation already captures the essential graph-theoretic physics, while allowing much larger numerical simulations.


This viewpoint connects the present problem to sparse SYK models, SYK chains, and network-based systems in which connectivity is an important dynamical variable.~\cite{GarciaGarciaJiaRosaVerbaarschot2021,TezukaOktayRinaldiHanadaNori2023,GuLucasQi2017,SongJianBalents2017,AltlandBagretsKamenev2019Granular}
In our case, however, the sparsity is not imposed by hand, it arises microscopically from overlap constraints. In this section, we make this nonuniform interaction structure explicit and quantify how it leads to the nucleation, growth, and eventual merger of emergent SYK clusters, namely groups of localized orbitals that are strongly mixed with one another while remaining only weakly coupled to the rest of the system.

\subsection{Description of the emergent graph}

A convenient representation of the quartic term in \refe{syk} is obtained by viewing $J_{ij}^{kl}$ as a matrix element for pair scattering. 
The operator $c_i^\dagger c_j^\dagger c_k c_l$ annihilates a fermion pair $(k,l)$ and creates a pair $(i,j)$, so it naturally acts in the space of two-particle "channels" labeled by an unordered index pair. 
We therefore introduce the pair label
\begin{equation}
p\equiv (i,j),\qquad 1\le i<j\le N,
\end{equation}
and rewrite the interaction tensor as a matrix $J_{pq}$ with $p=(i,j)$ and $q=(k,l)$, i.e.
\begin{equation}
J_{pq}\equiv J_{ij}^{kl}.
\end{equation}
The vertex set of the corresponding graph is the set of all pair labels,
\begin{eqnarray}
&&V \equiv \{\,p=(i,j)\,|\,1\le i<j\le N\,\},\nonumber\\
&&|V| \equiv N_{\rm pair}=\binom{N}{2}\sim \frac{N^2}{2}.
\end{eqnarray}
Hermiticity, $J_{ij}^{kl}=J_{kl}^{ij\,*}$, implies $J_{pq}=J_{qp}^*$, so the magnitude $|J_{pq}|$ is symmetric under $p\leftrightarrow q$. 
In what follows, we focus on this magnitude because it controls (i) whether a given pair channel is coupled at all (vanishing vs. nonvanishing matrix elements) and (ii) the hierarchy between "strong" and "weak" couplings in a finite sample (Sec.~\ref{sec:distJ}). 
We thus define a weighted, undirected graph $G$ on $V$ by connecting two distinct vertices $p\neq q$ whenever $J_{pq}\neq 0$, with positive edge weight
\begin{equation}
w_{pq}\equiv 
|J_{pq}|=|J_{ij}^{kl}|.
\end{equation}
Here $w_{pq}$ denotes the \emph{graph edge weight}, associated with four compact localized states (and should not be confused with the orbital widths $w_i$).

\begin{figure*}[t]
    \centering
    \includegraphics[width=.8\textwidth]{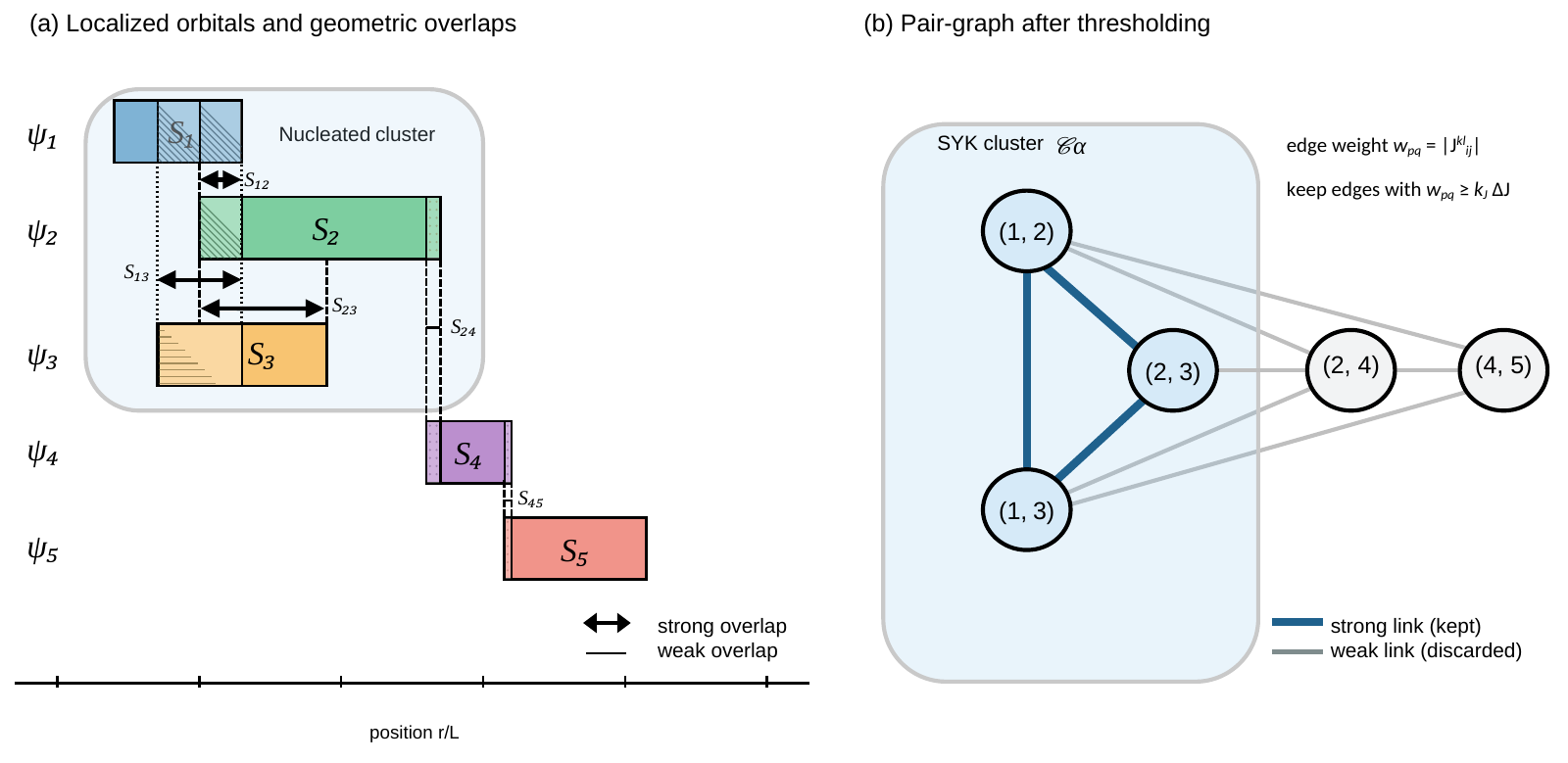}
    \caption{\capj{Schematic visualization of the interaction-graph construction used to diagnose SYK-cluster nucleation. 
    (a) A disorder realization of localized orbitals $\{\psi_i\}$ (Sec.~\ref{sec:geometric_construction}) generates quartic matrix elements $J_{ij}^{kl}$ via the overlap-dependent rules derived in Sec.~\ref{sec:correlated_disorder}. 
    Because many overlaps are empty, a finite fraction of couplings vanish exactly (point mass $P_0$ in \refe{PJA}), while the remaining magnitudes are broadly distributed (Sec.~\ref{sec:distJ}). 
    (b) The same data can be represented as a weighted graph in pair space: vertices are pair channels $p=(i,j)$, edges carry weights $w_{pq}=|J_{pq}|$, and a ``strong-link'' truncation keeps only edges with $w_{pq}\ge k_J\Delta J$. 
    Each connected component $\mathcal{C}_\alpha$ of the truncated graph defines an emergent SYK cluster; the associated orbital set $\mathcal{O}_\alpha$ is obtained by collecting all single-particle labels that appear in the vertices of $\mathcal{C}_\alpha$.}}
    \label{fig:pair_graph_schematic}
\end{figure*}

We ignore the diagonal elements $J_{pp}$, which correspond to self-loops in pair space and do not affect graph connectivity.
In this representation, a path $p\to q\to\cdots$ corresponds to a sequence of interaction-induced pair-scattering processes, so connected components encode which two-particle channels can be mixed without invoking parametrically small couplings.
A schematic of the mapping from localized orbitals to the pair-space interaction graph is shown in Fig.~\ref{fig:pair_graph_schematic}.

To isolate the interaction backbone that is most relevant for cluster nucleation, we introduce a strong-link truncation of the weighted graph. 
We first define a typical coupling scale by the root-mean-square (RMS) magnitude of off-diagonal pair-scattering matrix elements,
\begin{equation}
\Delta J \equiv 
\sqrt{\Big\langle |J_{pq}|^2 \Big\rangle_{p<q}}
\;=\;
\sqrt{\frac{2}{N_{\rm pair}(N_{\rm pair}-1)}\sum_{p<q}|J_{pq}|^2},
\end{equation}
Here $N_{\rm pair}=\binom{N}{2}$ is the total number of pair channels $p=(i,j)$ with $i<j$. The condition $p<q$ denotes an arbitrary but fixed ordering of pair labels used only to avoid double counting, and the diagonal $p=q$ is excluded. 
(For large $N_{\rm pair}$, the distinction between $N_{\rm pair}(N_{\rm pair}-1)$ and $N_{\rm pair}^2$ is negligible; we keep the exact normalization for definiteness.)
We then retain only those edges whose magnitude exceeds a threshold set by $\Delta J$,
\begin{equation}
w_{pq}=|J_{pq}|\ge k_J\,\Delta J,
\end{equation}
and discard all weaker links. 
The dimensionless parameter $k_J$ therefore controls the graph-theoretic coarse graining of the interaction: increasing $k_J$ zooms in on rarer, exceptionally strong couplings (resolving smaller, more weakly connected components), while decreasing $k_J$ progressively reincorporates weaker hybridization channels and eventually produces a single "giant" connected component. 
Unless stated otherwise, the numerical results in the next subsection use $k_J=1$ as a natural benchmark, keeping couplings above the RMS scale.

For the regularized Coulomb interaction $V_c(r)=1/\max(|r|,r_0)$, the parameter $r_0$ is a microscopic ultraviolet regulator. 
Once $r_0$ is chosen well below the typical overlap scale $\overline{w}L$, the distribution of induced couplings converges (Fig.~\ref{fig:J_hist}) and the topology of the truncated interaction graph (connected components, clique counts, scaling exponents) becomes insensitive to the precise value of $r_0$. 
In the numerical work, we therefore treat $r_0$ as a numerical cutoff rather than as a physical control parameter.

After truncation, the strong-link graph $G(k_J)$ decomposes into a disjoint union of connected components,
\begin{equation}
G(k_J)=\bigcup_\alpha \mathcal{C}_\alpha ,
\end{equation}
where each $\mathcal{C}_\alpha$ is a maximal connected subgraph in pair space. 
We interpret each $\mathcal{C}_\alpha$ as an \emph{emergent SYK cluster}: it is the set of pair channels that are mutually hybridized by a network of strong interaction matrix elements, and therefore forms an approximately self-contained subspace for interaction-driven dynamics on intermediate time/energy scales. 
To connect this notion back to the localized orbitals of Secs.~\ref{sec:correlated_disorder} and \ref{sec:true_syk}, using graph theoretic notations, we associate with each component the set of single-particle labels that appear in its vertices,
\begin{eqnarray}
&&\mathcal{O}_\alpha \equiv \Big\{\, i\in\{1,\dots , N\}\; \Big|\; \exists\, j\neq i \ \text{s.t.}\nonumber\\ 
&&\qquad\qquad\qquad\qquad\big(\min\{i,j\},\max\{i,j\}\big)\in \mathcal{C}_\alpha \Big\},\nonumber\\
&&N^{\rm orb}_\alpha \equiv |\mathcal{O}_\alpha|.
\end{eqnarray}
Physically, $\mathcal{O}_\alpha$ is the subset of localized orbitals that have nucleated into a single strongly correlated cluster: repeated interaction events can scatter fermion pairs within $\mathcal{O}_\alpha$ without requiring weak links to orbitals outside. 
When $\mathcal{C}_\alpha$ is sufficiently dense that it contains (most of) the pair channels formed from $\mathcal{O}_\alpha$, one expects $|\mathcal{C}_\alpha|\simeq \binom{N^{\rm orb}_\alpha}{2}$, so that $N^{\rm orb}_\alpha$ provides a direct estimate of the effective SYK cluster size discussed in the Introduction. 
Even when $\mathcal{C}_\alpha$ is sparse in pair space, $N^{\rm orb}_\alpha$ remains the most physically transparent measure of how many localized degrees of freedom participate in the cluster.

To quantify the internal connectivity of each cluster, we introduce a hierarchy of simplex (clique) numbers. 
Let $N^{(0)}_\alpha$ be the number of vertices (pair channels) in $\mathcal{C}_\alpha$ and $N^{(1)}_\alpha$ the number of retained edges (strong couplings) inside $\mathcal{C}_\alpha$. 
More generally, we define $N^{(n)}_\alpha$ to be the number of \emph{$n$-simplexes} in $\mathcal{C}_\alpha$, i.e.\ the number of complete subgraphs on $n{+}1$ vertices. 
Thus $N^{(2)}_\alpha$ counts triangles (3-cliques), $N^{(3)}_\alpha$ counts tetrahedra (4-cliques), and so on. 
These higher-order counts are a direct diagnostic of "SYK-likeness": a dense, all-to-all cluster contains parametrically many short loops and high-order cliques, whereas a sparse, geometry-constrained cluster is closer to a locally tree-like network.

A compact way to summarize how $N^{(n)}_\alpha$ scales with cluster size is via the exponents
\begin{equation}
C^{(n)}_\alpha \equiv \frac{\ln N^{(n)}_\alpha}{\ln N^{(0)}_\alpha},
\end{equation}
so that
\begin{equation}
N^{(n)}_\alpha \sim \big[N^{(0)}_\alpha\big]^{\,C^{(n)}_\alpha},
\end{equation}
when $C^{(n)}_\alpha$ is weakly size dependent.
In the next Sec.~IV\,B we will visualize these scalings via log-log plots of $N^{(n)}_\alpha$ versus $N^{(n-1)}_\alpha$ (rather than always versus $N^{(0)}_\alpha$). In that representation, the fitted slope equals $C^{(n)}_\alpha/C^{(n-1)}_\alpha$ for $n\ge2$. For a complete graph one has $C^{(1)}=2$, $C^{(2)}=3$, $C^{(3)}=4$, hence the corresponding slopes are $3/2$ and $4/3$ (the complete-graph benchmarks).

The case $n=1$ distinguishes sparse from dense connectivity. 
If the average degree
\begin{equation}
D_\alpha \equiv \frac{2N^{(1)}_\alpha}{N^{(0)}_\alpha}
\end{equation}
approaches a constant as $N^{(0)}_\alpha$ grows, then $N^{(1)}_\alpha\propto N^{(0)}_\alpha$ and $C^{(1)}_\alpha\to 1$, as in graphs with an effectively finite coordination number. 
At the opposite extreme, one may define the standard edge density
\begin{equation}
\rho_\alpha \equiv \frac{N^{(1)}_\alpha}{\binom{N^{(0)}_\alpha}{2}}
=\frac{2N^{(1)}_\alpha}{N^{(0)}_\alpha\big(N^{(0)}_\alpha-1\big)}.
\label{eq:edge_dense}
\end{equation}
If $\rho_\alpha$ approaches a nonzero constant, then $N^{(1)}_\alpha\propto [N^{(0)}_\alpha]^2$ and $C^{(1)}_\alpha\to 2$, while higher-order simplex counts scale as $C^{(n)}_\alpha\to n{+}1$ as in a complete graph. 
This dense scaling is the closest network-theoretic proxy for the fully connected interaction pattern of the canonical SYK ensemble. 
Intermediate values $1<C^{(1)}_\alpha<2$ correspond to clusters that are neither locally tree-like nor fully connected. A useful example is a rectangular graph in which the sites are connected very differently along the two directions: along one direction, every site is connected to all others, while along the other direction, each site is connected only to its nearest neighbors. If both sides of the rectangle have length $L$, then the number of vertices scales as $N^{(0)}_\alpha\sim L^2,$ whereas the number of higher-order simplexes scales as $N^{(k)}_\alpha\sim L^{k+2}$. This gives $C^{(k)}_\alpha=1+k/2$, and in particular $C^{(1)}_\alpha=3/2$. The numerical results below show that the width disorder discussed in \refs{width_disorder} provides a microscopic way to tune the effective interaction graph between sparse and dense regimes, by controlling how many couplings are exactly zero and how many remain large.

\subsection{Numerical analysis of the emergent graph}

We now quantify this graph structure directly. The basic observables are the connected components $\{\mathcal{C}_\alpha\}$ of the strong-link truncated pair graph $G(k_J)$ introduced above. Pair channels are labeled by $p=(i,j)$ and $q=(k,l)$, edge weights are $w_{pq}=|J_{pq}|$, and only links satisfying the threshold $|J_{pq}|\ge k_J\Delta J$ are retained, with $\Delta J$ the RMS coupling scale. The resulting components are the candidate SYK clusters generated by the microscopic overlap geometry.

For each population size $N$, we sample a disorder realization of localized orbitals in a system of length $L=1$ and compare two width ensembles: variable widths $w_i\sim{\rm Unif}[0,w]$ and constant width $w_i=w$. Unless stated otherwise, we use the benchmark threshold $k_J=1$ and a short-distance regulator $r_0\ll w$ chosen well inside the converged regime of Fig.~\ref{fig:J_hist}, so that the graph observables are insensitive to the precise ultraviolet cutoff. From the induced interaction tensor we construct the weighted pair graph on $N_{\rm pair}=\binom{N}{2}$ vertices and compute its connected components $\mathcal{C}_\alpha$, ordered by decreasing vertex count $N^{(0)}_\alpha\equiv|\mathcal{C}_\alpha|$. For each component we also evaluate the simplex numbers introduced in Sec.~IV A: $N^{(1)}_\alpha$ counts retained edges, $N^{(2)}_\alpha$ triangles, and $N^{(3)}_\alpha$ tetrahedra. Power-law fits of the form $N^{(n)}_\alpha\propto\left(N^{(0)}_\alpha\right)^{C^{(n)}_\alpha}$ then quantify how close a given component is to sparse, intermediate, or nearly complete-graph behavior.


\begin{figure*}
    \centering
    \begin{subfigure}[l]{\linewidth}
        \centering
        \includegraphics[width=.40\textwidth]{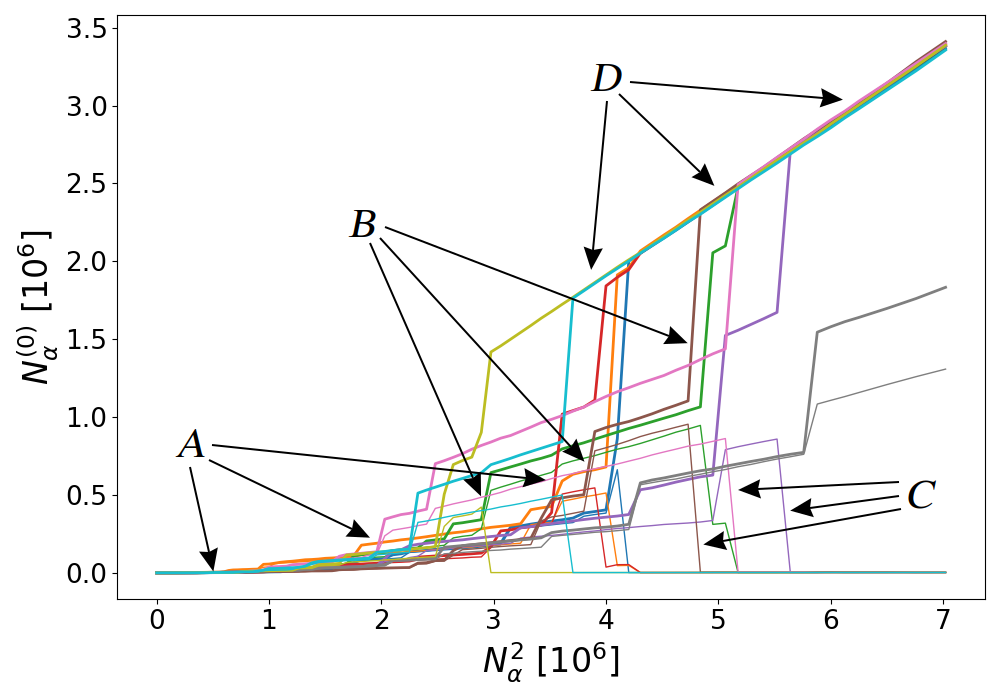}
        \includegraphics[width=.40\textwidth]{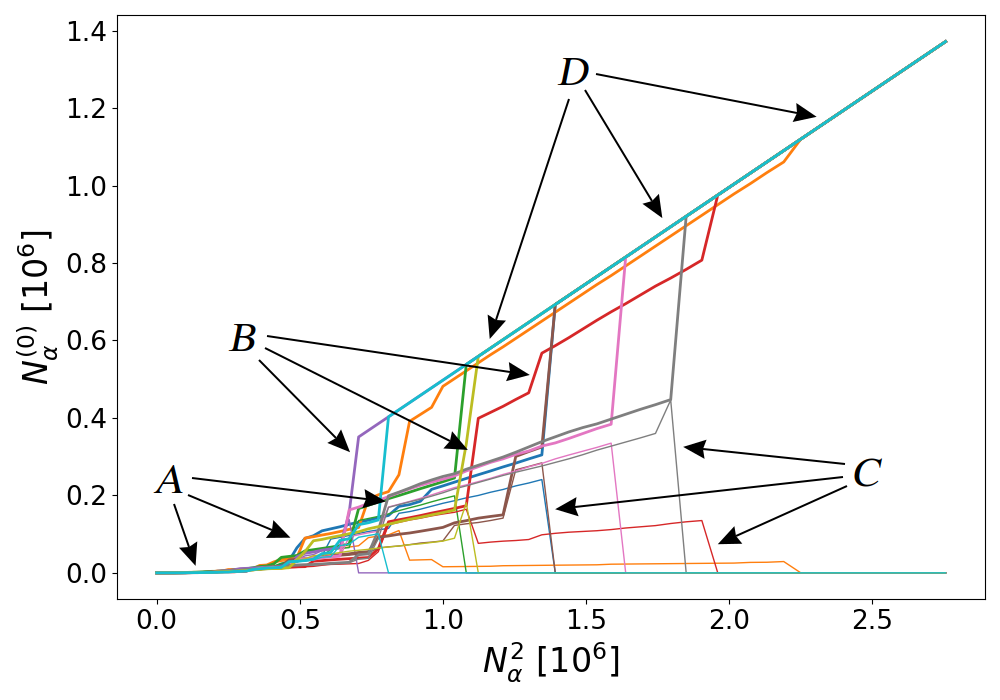}
        \caption{\capj{Cluster sizes $N^{(0)}_\alpha$ on the squared population $N^2$.}}
        \label{fig:size_on_N-1}
    \end{subfigure}
    \begin{subfigure}[l]{\linewidth}
        \centering
        \includegraphics[width=.40\textwidth]{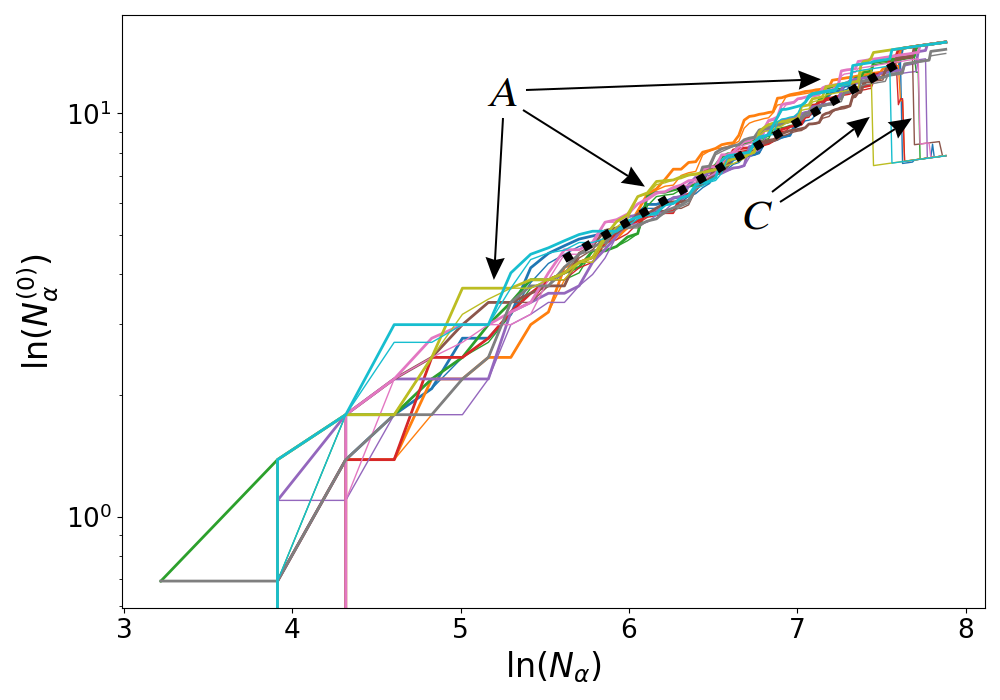}
        \includegraphics[width=.40\textwidth]{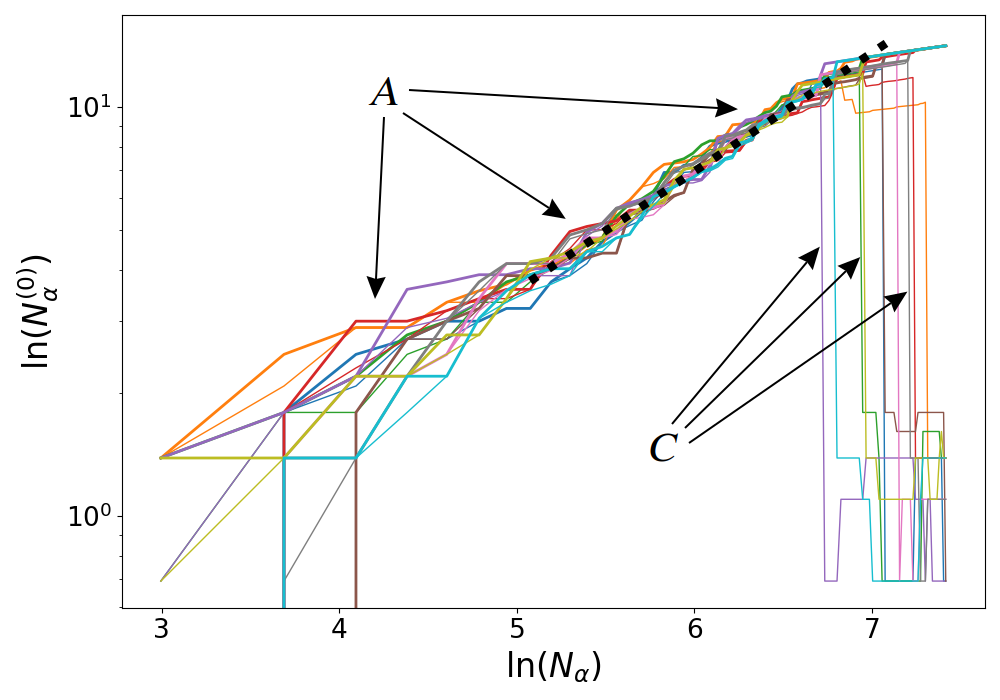}
        \caption{\capj{The same data on logarithmic axes, together with stretched-exponential fits in the pre-saturation regime.}}
        \label{fig:size_on_N-2}
    \end{subfigure}
    \caption{\capj{Connected-component sizes in the strong-link truncated pair graph $G(k_J)$. For each disorder realization (color) and population size $N$, we construct the graph with vertices $p=(i,j)$, retain only edges with $|J_{pq}|\ge k_J\Delta J$ (here $k_J=1$), and order the connected components $\{\mathcal{C}_\alpha\}$ by decreasing vertex count $N^{(0)}_\alpha=|\mathcal{C}_\alpha|$. Thick curves denote the largest component and thin curves several subleading components. The left column shows the variable-width ensemble $w_i\sim{\rm Unif}[0,w]$ and the right column the constant-width ensemble $w_i=w$, with $w=0.01$ and all lengths measured in units of $L$. Panels (a) plot $N^{(0)}_\alpha$ versus $N^2$, the natural scaling variable of the full pair space $N_{\rm pair}\sim N^2/2$. The labels $A$-$D$ mark four recurrent stages: nucleation and irregular early growth, component mergers, onset of the saturated/percolated regime, and linear-in-$N^2$ growth of the giant component. Panels (b) show the same data on logarithmic axes together with stretched-exponential fits $N^{(0)}_\alpha\propto \exp(\beta N^\gamma)$ (dashed black) used to characterize the pre-saturation regime.}}
    \label{fig:dot_groth}
\end{figure*}



\begin{figure*}
    \begin{subfigure}[t]{0.74\linewidth}
        \centering
        \includegraphics[width=.30\textwidth]{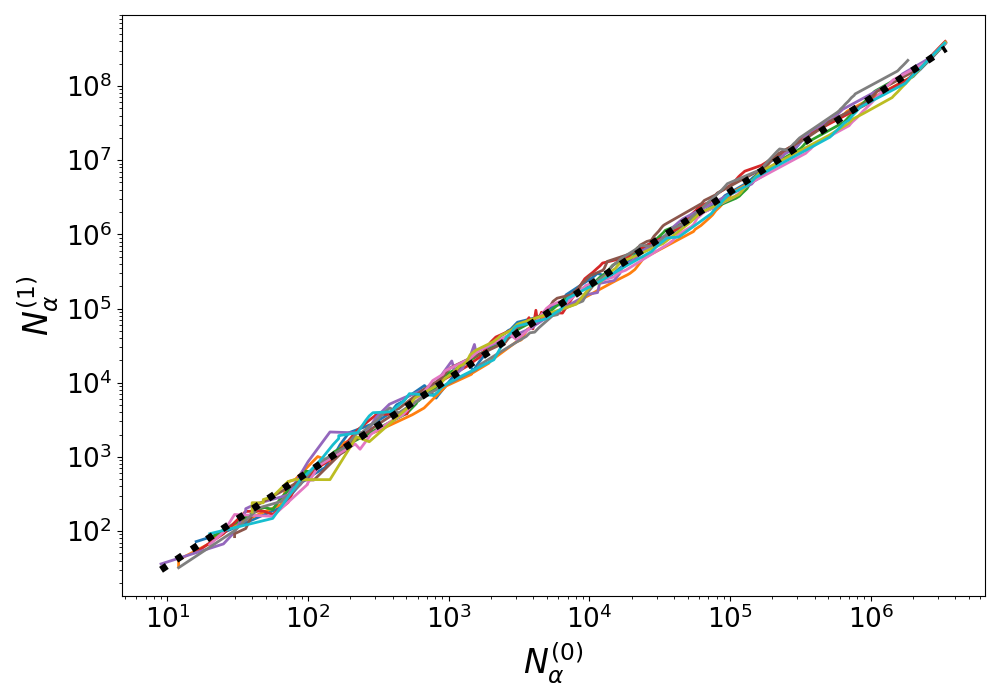}\hfill
        \includegraphics[width=.30\textwidth]{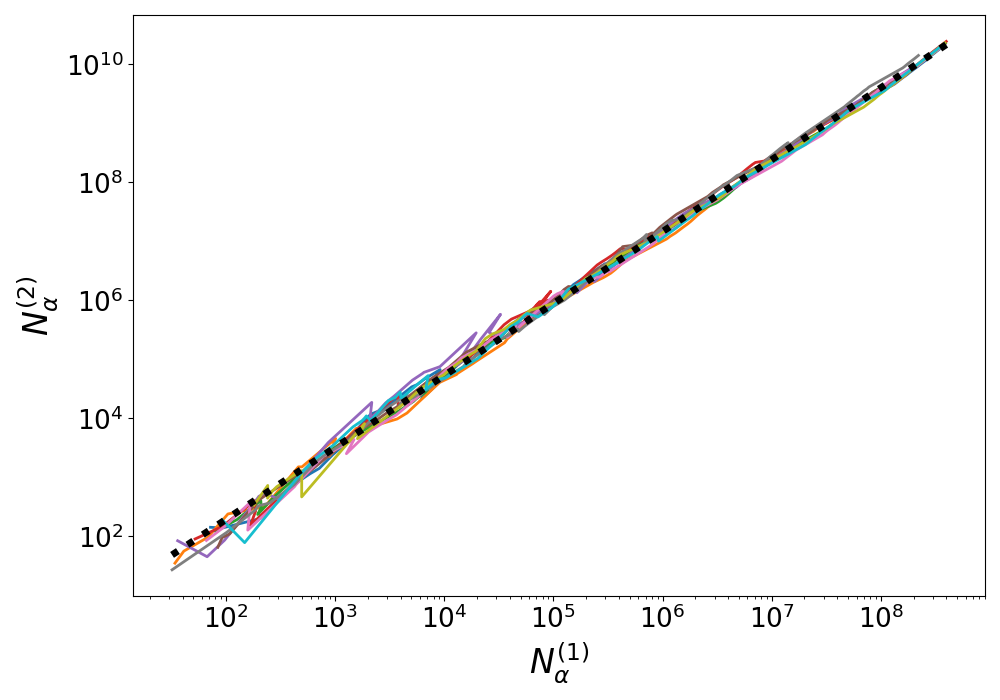}\hfill
        \includegraphics[width=.30\textwidth]{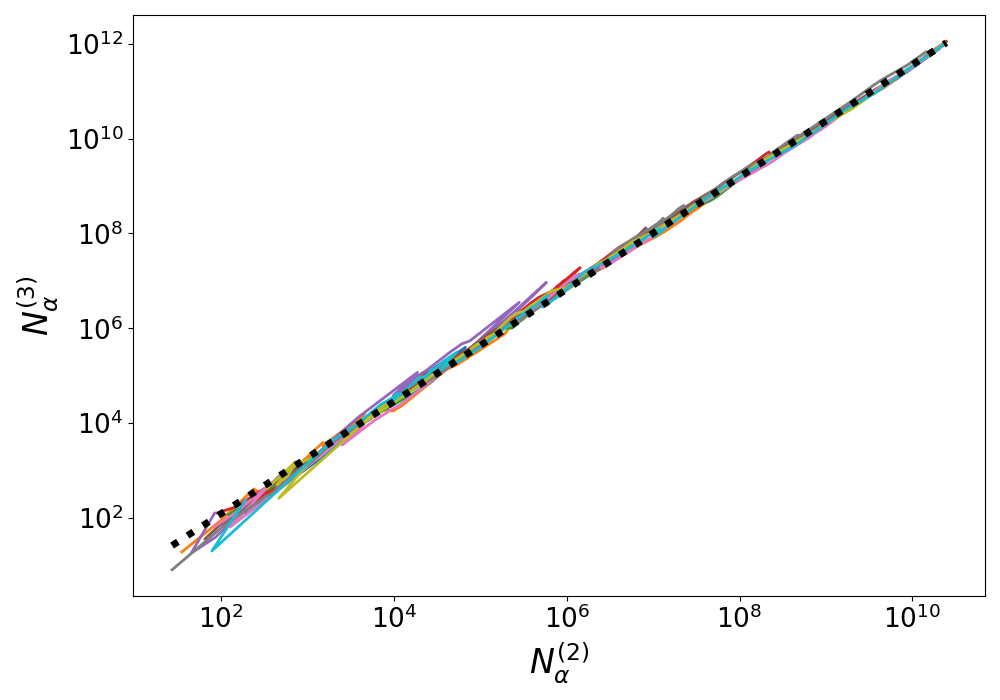}
        \caption{\capj{$w_i\in[0,\overline{w}]$, the non-saturated regime.}}
        \label{fig:Na_on_N0_vw}
    \end{subfigure}
    \vspace{0.35em}
    \begin{subfigure}[t]{0.74\linewidth}
        \centering
        \includegraphics[width=.30\textwidth]{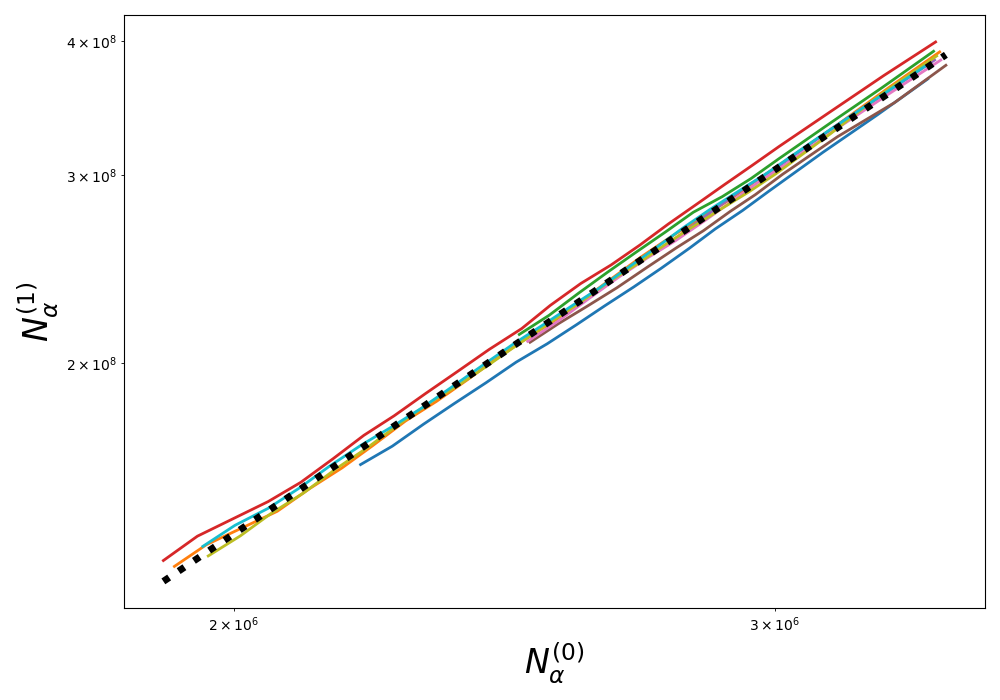}\hfill
        \includegraphics[width=.30\textwidth]{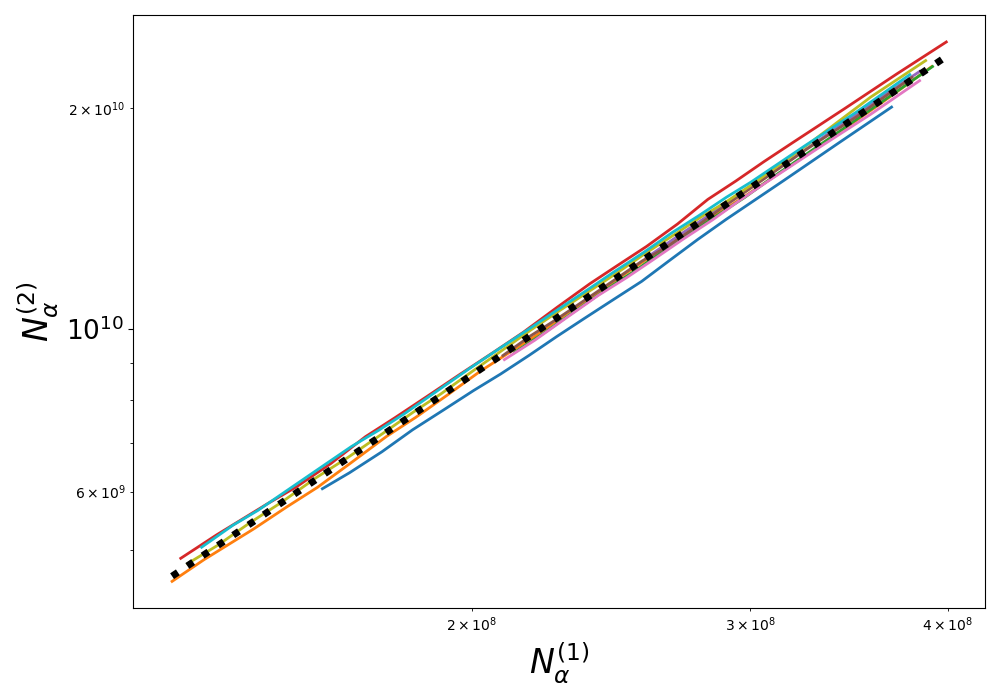}\hfill
        \includegraphics[width=.30\textwidth]{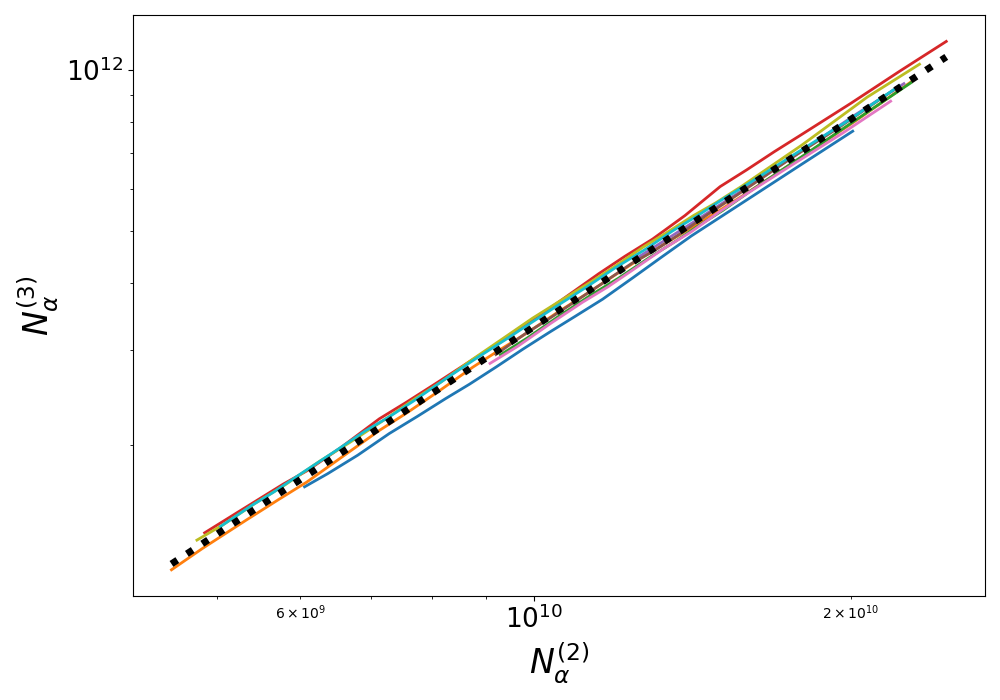}
        \caption{\capj{$w_i\in[0,\overline{w}]$, the saturated regime.}}
        \label{fig:Na_on_N0_vw_sat}
    \end{subfigure}
    \vspace{0.35em}
    \begin{subfigure}[t]{0.74\linewidth}
        \centering
        \includegraphics[width=.30\textwidth]{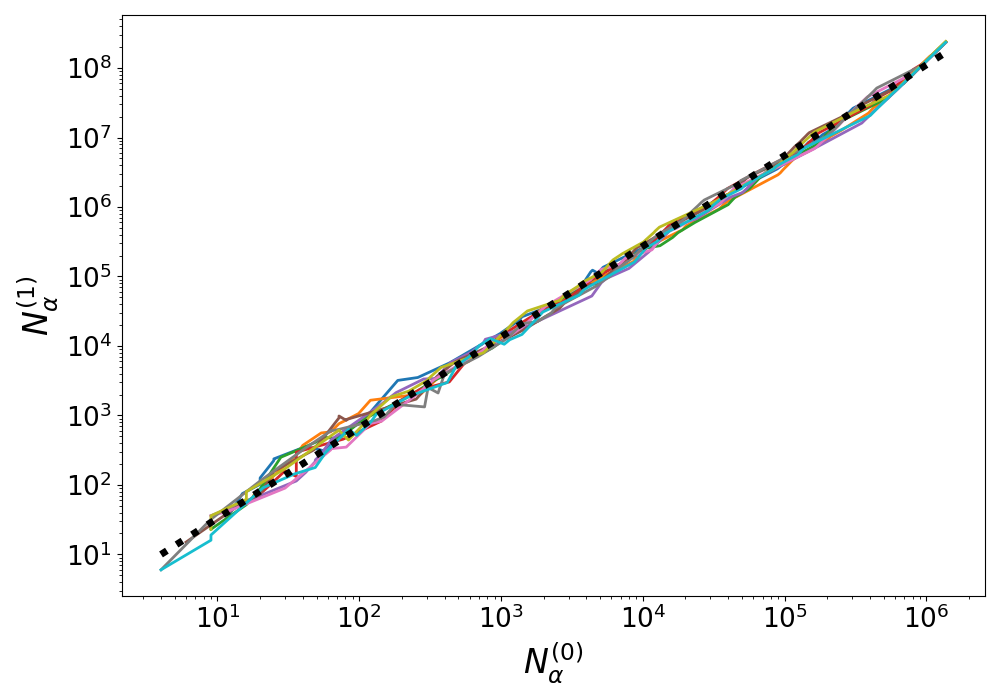}\hfill
        \includegraphics[width=.30\textwidth]{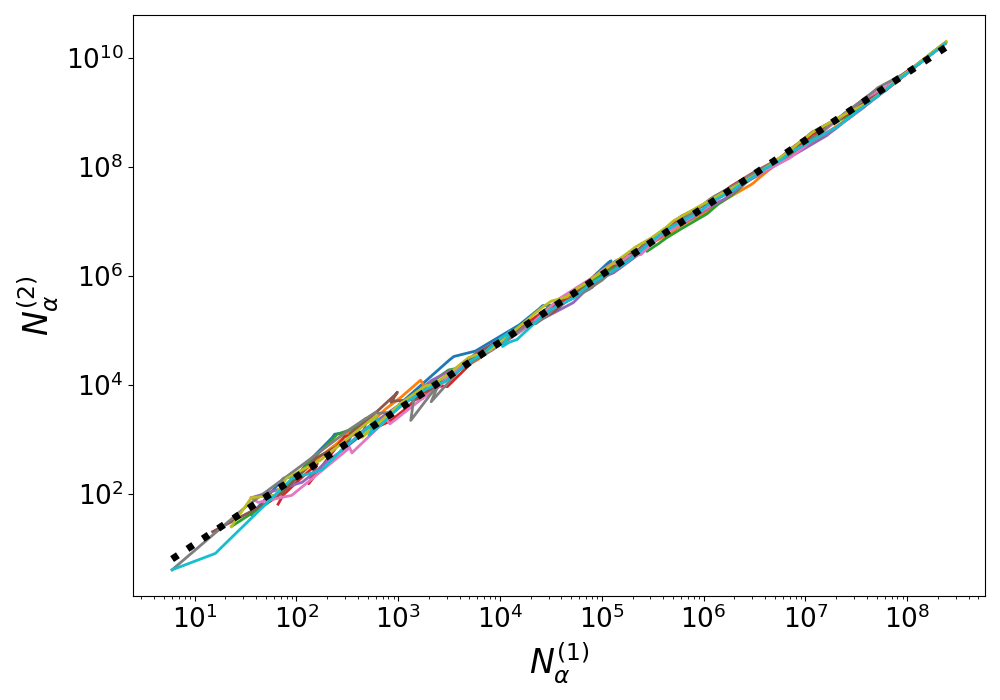}\hfill
        \includegraphics[width=.30\textwidth]{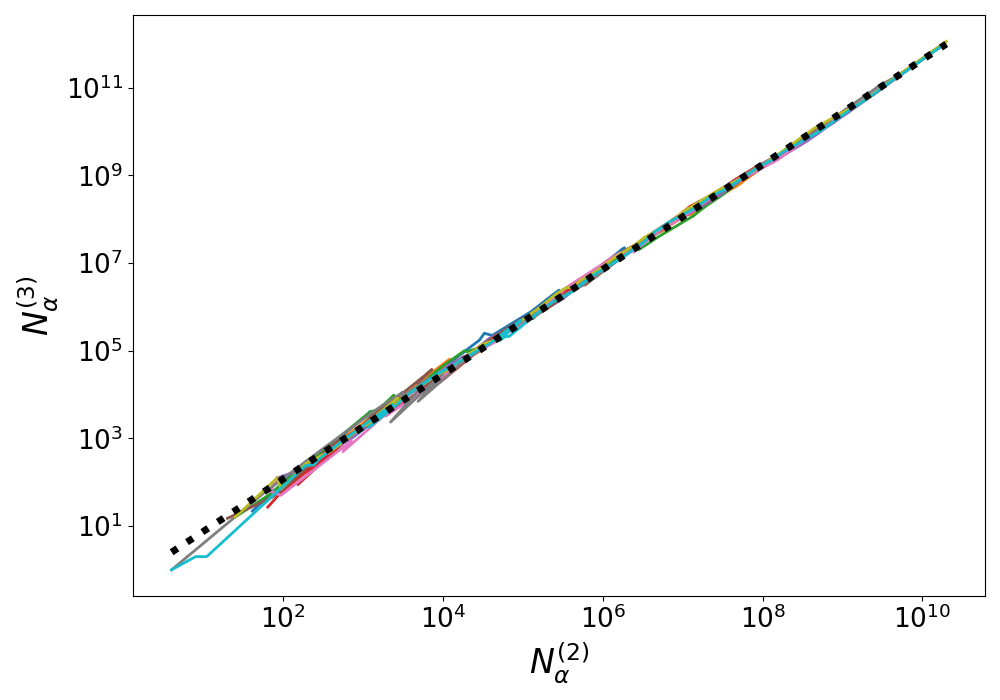}
        \caption{\capj{$w_i=\overline{w}$, the non-saturated regime.}}
        \label{fig:Na_on_N0_cw}
    \end{subfigure}
    \vspace{0.35em}
    \begin{subfigure}[t]{0.74\linewidth}
        \centering
        \includegraphics[width=.30\textwidth]{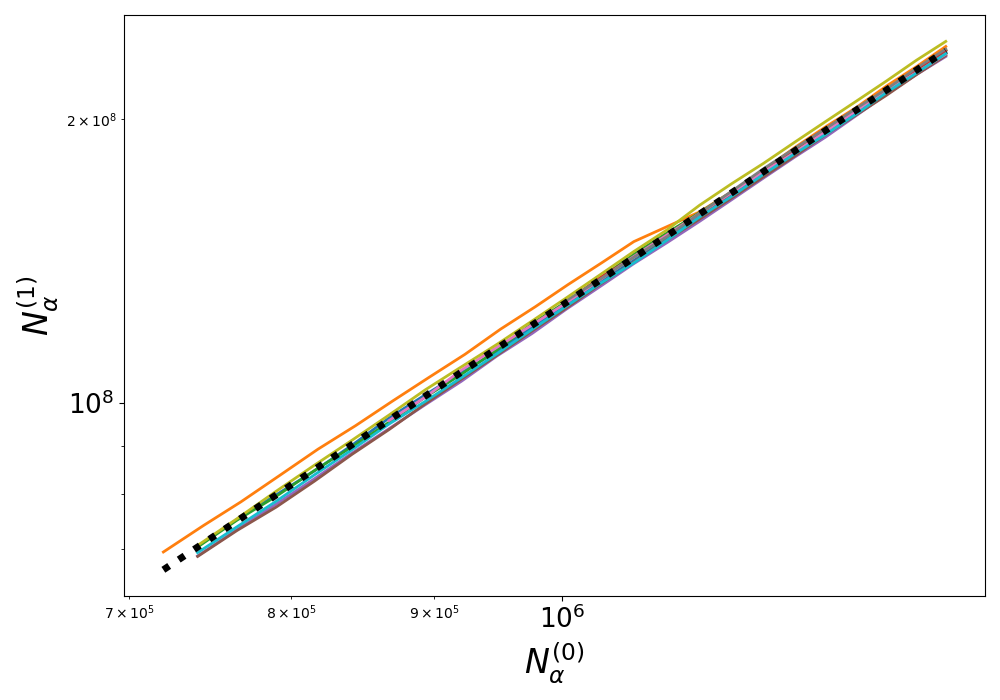}\hfill
        \includegraphics[width=.30\textwidth]{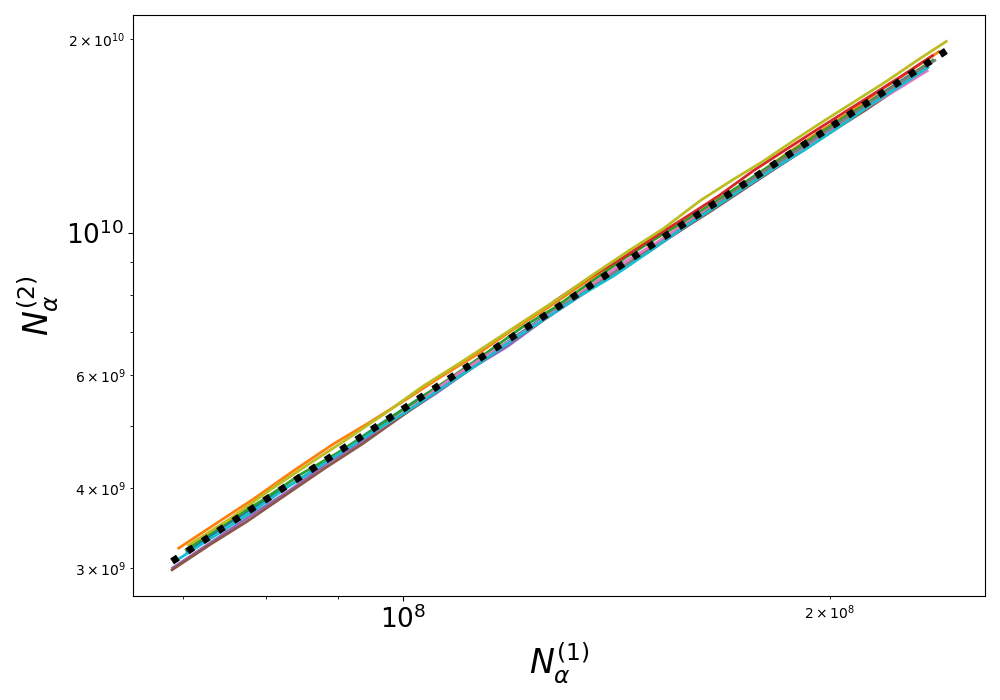}\hfill
        \includegraphics[width=.30\textwidth]{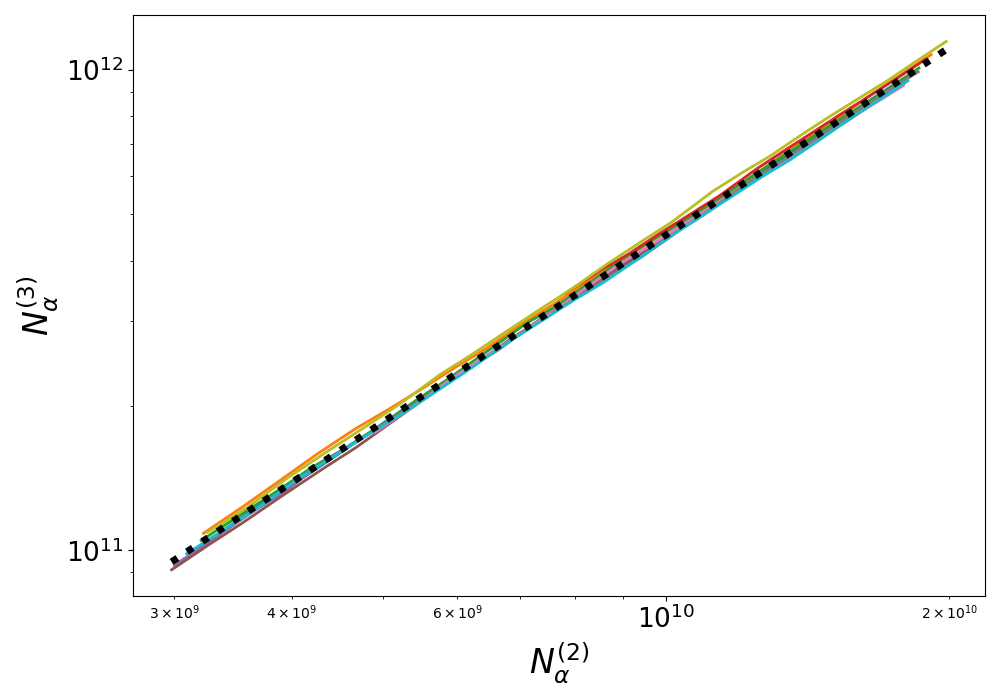}
        \caption{\capj{$w_i=\overline{w}$, the saturated regime.}}
        \label{fig:Na_on_N0_cw_sat}
    \end{subfigure}
    \caption{\capj{Clique (simplex) scaling inside the largest connected components of the truncated pair graph $G(k_J)$, for the same parameters as \reff{dot_groth} ($k_J=1$, $w=0.01$). For each component $\mathcal{C}_\alpha$ we compute the simplex numbers $N^{(n)}_\alpha$ introduced in Sec.~IV A: $N^{(0)}_\alpha$ vertices, $N^{(1)}_\alpha$ retained edges, $N^{(2)}_\alpha$ triangles, and $N^{(3)}_\alpha$ tetrahedra. Each panel shows multiple disorder realizations (colors) together with power-law fits (dashed black). The three columns plot $N^{(1)}_\alpha$ vs.\ $N^{(0)}_\alpha$, $N^{(2)}_\alpha$ vs.\ $N^{(1)}_\alpha$, and $N^{(3)}_\alpha$ vs.\ $N^{(2)}_\alpha$ on log--log axes; when $N^{(n)}_\alpha\propto (N^{(0)}_\alpha)^{C^{(n)}_\alpha}$, the corresponding slopes are $C^{(1)}_\alpha$, $C^{(2)}_\alpha/C^{(1)}_\alpha$, and $C^{(3)}_\alpha/C^{(2)}_\alpha$. Panels (a,b) correspond to the variable-width ensemble $w_i\in[0,w]$ and panels (c,d) to constant width $w_i=w$. Panels (a,c) show the non-saturated regime and panels (b,d) the saturated regime. The systematic drift of the fitted slopes toward the complete-graph benchmarks $(C^{(1)},C^{(2)},C^{(3)})=(2,3,4)$ in panels (b,d) shows that the giant component becomes increasingly dense in pair space, with the constant-width ensemble consistently closer to the complete-graph limit.}}
    \label{fig:dot_scaling}
\end{figure*}


In \reff{dot_groth} we show the sizes $N^{(0)}_\alpha$ of the largest connected component (thick line) and several subleading components (thin lines) as the population $N$ is increased, for multiple disorder realizations (colors), and for both width ensembles (left: variable widths; right: constant width).
Because the vertex set itself scales as $N_{\rm pair}\sim N^2/2$, panel~(a) plots $N^{(0)}_\alpha$ against $N^2$ to make the extensive growth of a giant component visually transparent.
Four recurrent qualitative features, indicated by the labels $A$-$D$ in \reff{dot_groth}(a), need to be emphasized:

$A$: \emph{Nucleation and irregular early growth.}
At smaller $N$ the truncated graph is fragmented into many components.  The largest few components grow in a strongly realization-dependent, staircase-like fashion.
This is expected because increasing $N$ introduces (i) new pair vertices and (ii) many new candidate edges whose retention depends on the global scale $\Delta J$; as a result, cluster growth is not self-averaging at moderate $N$.

$B$: \emph{Mergers (vertical jumps)}.
Sharp upward steps correspond to the merger of two previously disconnected components when a newly introduced strong edge (or a short chain of strong edges) bridges them.
These jumps are the graph-theoretic signature of "cluster coalescence" and are the microscopic origin of the non-smooth component-size curves.

$C$: \emph{Onset of the saturated (percolated) regime}.
In a realization-dependent population, the largest component becomes parametrically larger than the second-largest one, i.e., the component-size hierarchy develops a clear gap.
We use this qualitative criterion to identify the transition into the saturated regime.

$D$: \emph{Linear-in-$N^2$ growth of the giant component.}
Deep in the saturated regime the largest component occupies essentially the entire available pair-vertex space, and its size scales as
$N^{(0)}_{\rm g}\propto N^2$ (panel~(a)), consistent with $N^{(0)}_{\rm g}\simeq N_{\rm pair}=\binom{N}{2}$ up to a $k_J$-dependent prefactor.

In \reff{dot_groth}(b), we show the same data on logarithmic axes to quantify the pre-saturation growth law.  In the non-saturated regime, the typical component size is well described by a stretched-exponential form
\begin{equation}
N^{(0)}_\alpha \propto \exp\!\left(\beta N^\gamma\right),
\end{equation}
with $\gamma_v\simeq 0.56$ for variable widths and $\gamma_c\simeq 2/3$ for constant width (black dashed fits).
The larger value of $\gamma$ for the constant-width ensemble is consistent with the overlap statistics derived in \refs{width_disorder}.  In particular, after setting $L=1$ so that $w\equiv \overline{w}$, \refee{overlap_distr_ecw}{overlap_P0_vw} imply $\mathrm{Prob}(\ell_{ij}>0)=1-(1-w)^2\simeq 2w$ for the constant-width ensemble, whereas $\mathrm{Prob}(\ell_{ij}>0)=1-\mathrm{Prob}(\ell_{ij}=0)=w-\tfrac{7}{24}w^2\simeq w$ for the variable-width ensemble.  This yields a parametrically larger density of nonzero (and hence potentially strong) matrix elements in the constant-width case and therefore a faster approach to percolation.

While \reff{dot_groth} diagnoses which pair channels become connected by strong couplings, \reff{dot_scaling} probes how densely they are connected inside each component through the simplex numbers $N^{(n)}_\alpha$.
The four panels separate $(a,c)$ the non-saturated regime from $(b,d)$ the saturated regime, and compare $(a,b)$ variable widths to $(c,d)$ constant width.
Within each panel, the three columns show log-log plots of $N^{(1)}_\alpha$ vs.\ $N^{(0)}_\alpha$ (left), $N^{(2)}_\alpha$ vs.\ $N^{(1)}_\alpha$ (middle), and $N^{(3)}_\alpha$ vs.\ $N^{(2)}_\alpha$ (right); dashed black lines are power-law fits.
For reference, a complete graph has $C^{(1)}=2$, $C^{(2)}=3$, $C^{(3)}=4$, corresponding to slopes $C^{(2)}/C^{(1)}=3/2$ and $C^{(3)}/C^{(2)}=4/3$. 

Two quantitative conclusions follow directly from the fitted exponents quoted below.

1) \emph{Non-saturated clusters are sparse but increasingly loopy.} In the non-saturated regime we find 
\begin{equation}
(C^{(1)},C^{(2)},C^{(3)}) \approx (1.26,1.53,1.81)\quad\text{for }w_i\in[0,\overline w],\nonumber
\end{equation}
as seen from \reff{Na_on_N0_vw}), and 
\begin{equation}
(C^{(1)},C^{(2)},C^{(3)}) \approx (1.31,1.61,1.92)\quad\text{for }w_i=\overline w,\nonumber
\end{equation}
as seen from \reff{Na_on_N0_cw}. Since $C^{(1)}<2$, the retained strong-link subgraph is \emph{sub-dense}: using the standard edge density \refe{edge_dense}, one has $\rho_\alpha\sim [N^{(0)}_\alpha]^{C^{(1)}-2}$, so the density of retained strong links decreases with component size in this regime.
At the same time, $C^{(2)}>C^{(1)}$ and $C^{(3)}>C^{(2)}$ indicate a growing abundance of short loops and higher-order cliques as the component expands, i.e. a progressive crossover away from a locally tree-like network.

2) \emph{Saturated clusters approach complete-graph (SYK-like) scaling.}
In the saturated regime, the exponents become much closer to the complete-graph benchmarks:
\begin{equation}
(C^{(1)}_{\rm s},C^{(2)}_{\rm s},C^{(3)}_{\rm s}) \approx (1.93,2.78,3.55)\quad\text{for }w_i\in[0,\overline w],\nonumber
\end{equation}
as seen from \reff{Na_on_N0_vw_sat}, and 
\begin{equation}
(C^{(1)}_{\rm s},C^{(2)}_{\rm s},C^{(3)}_{\rm s}) \approx (1.97,2.86,3.70)\quad\text{for }w_i=\overline w,\nonumber
\end{equation}
as seen from \reff{Na_on_N0_cw_sat}. Equivalently, the fitted slopes in \reff{dot_scaling} are close to the complete-graph values:
$C^{(2)}_{\rm s}/C^{(1)}_{\rm s}\simeq 1.44$--$1.45$ (vs. $3/2$) and
$C^{(3)}_{\rm s}/C^{(2)}_{\rm s}\simeq 1.28$--$1.29$ (vs. $4/3$).
Thus, once the giant component forms, the strong-link backbone inside it becomes almost maximally dense in the scaling sense and supports parametrically many triangles and tetrahedra, a key network-theoretic proxy for SYK-like all-to-all mixing in pair space.
The \emph{constant-width ensemble} remains systematically closer to the complete-graph values, consistent with reduced geometric sparsity relative to the variable-width ensemble.

Even when $C^{(1)}_{\rm s}\approx 2$, the prefactor of the edge scaling matters because $k_J$-thresholding retains only the strongest couplings.  Relative to the standard density measure \refe{edge_dense}, a convenient way to parametrize this is to write
$N^{(1)}_{\rm g}\simeq \tilde\rho_{\rm s}\,\left(N^{(0)}_{\rm g}\right)^{C^{(1)}_{\rm s}}$
for the giant component and interpret $\tilde\rho_{\rm s}$ as a generalized (scale-invariant) strong-link density.
For the parameters of \reff{dot_groth} and \reff{dot_scaling} ($k_J=1$, $\overline w=0.01$) we obtain
$\tilde\rho_{\rm s}\approx 5\times 10^{-5}$ for $w_i\in[0,\overline w]$ and
$\tilde\rho_{\rm s}\approx 10^{-4}$ for $w_i=\overline w$.
The factor-of-two enhancement for \emph{constant width} again reflects the higher prevalence of overlapping orbitals and, correspondingly, of large-amplitude couplings that survive the strong-link truncation.


\section{Conclusions}

We developed a minimal real-space theory of how SYK-like interaction clusters nucleate from localized single-particle states on an effectively one-dimensional manifold. The analysis has three steps. First, projecting a local interaction onto coarse localized seed envelopes produces a correlated disorder ensemble rather than the canonical SYK one: the couplings are rotationally invariant in the complex plane, but they retain a finite point mass at zero, a broad non-Gaussian continuous sector, and strong inter-coupling correlations inherited from common overlap geometry. This is the natural microscopic baseline for condensed-matter realizations, where the couplings are generated by real-space structure.

Second, we showed how the active sector can approach the complex-SYK Gaussian law. When each localization volume contains many microscopic pieces with independently randomized phases, a typical nonzero matrix element becomes a sum of many small complex contributions, and the active sector Gaussianizes. The random partition ensemble and the equal-cell partition ensemble both approach the same large-$M$ limit, showing that complete microscopic independence is unnecessary. The random partition mainly changes the finite-$M$ crossover and the near-zero sector. The zero pattern set by parent overlap remains present. The large-$M$ system is therefore a sparse, asymptotically canonical SYK interaction tensor, not a single fully connected SYK dot.

Third, we mapped this tensor to a graph in pair space and used connected components and simplex counts to describe cluster nucleation. The numerics show a clear sequence of nucleation, mergers, and saturation, and the simplex exponents drift toward complete-graph benchmarks as the giant component forms. In this language, approaching SYK physics is not only observed from the one-point distribution of couplings, it is also seen from how densely the strong links fill pair space.


The physical degrees of freedom are canonical fermion operators constructed from orthonormal localized orbitals. Spatial overlap between these orbitals is allowed, and actually necessary for the interaction matrix elements to be nonzero. In the phase-random large-$M$ regime, the overlap matrix becomes close to the identity, so orthogonalization is controlled. In the limit $M\to\infty$, this orthogonalized theory reaches the exact Gaussian one-point law in the active sector, while connected correlations vanish for generic distinct couplings built from disjoint quartets. The residual correlations are small in $1/N^2$, and are confined to a non-generic subset of pairs sharing an entire microscopic leg. Combining these observations, it becomes clear that in the limits $M\rightarrow\infty$ and the thermodynamic limit $N\rightarrow \infty$, this orthogonalized thermodynamic theory reaches the exact SYK disorder law for the active sector.

These results translate into concrete microscopic criteria. A promising platform should provide (i) an underlying canonical fermion field, (ii) an effectively one-dimensional set of localized modes, such as an irregular boundary, edge, or filament, (iii) sufficient real-space overlap between neighboring orthonormal orbitals, (iv) broken reality of the internal wave function through magnetic field, complex hopping, or channel mixing, and (v) enough internal phase-randomized structure that a typical active quartet has $M_{\rm eff}\gg1$. A natural microscopic description is therefore not a strictly one-channel real 1D Schr\"odinger Hamiltonian, but a multicomponent or higher-dimensional parent Hamiltonian whose low-energy states are localized along an effectively one-dimensional manifold. 
Under those conditions one expects a crossover from isolated SYK droplets to a giant interacting component, with corresponding consequences for sample-to-sample fluctuations, tunneling spectra, noise, and transport~\cite{CanNicaFranz2019,BrzezinskaGuanYazyevSachdevKruchkov2023,AltlandBagretsKamenev2019Transport,PavlovKiselev2026}. More broadly, the present work identifies the minimal real-space ingredients for SYK-cluster nucleation and provides a simple language -Gaussianization of the active sector, graph percolation, and simplex scaling - for analyzing how that limit is approached in realistic finite systems.

\section*{Acknowledgments}
The authors thank Hayk Mikayelyan and Roderich Moessner for helpful discussions. The research was supported by the Armenian Higher Education and Science Committee under the ARPI Remote Laboratory program 24RL-1C024.


\end{document}